\documentclass[12pt,preprint]{aastex}
\usepackage{psfig}

\newcommand{\beq}       {\begin{equation}}
\newcommand{\eeq}       {\end{equation}}
\newcommand{\beqa}      {\begin{eqnarray}}
\newcommand{\eeqa}      {\end{eqnarray}}

\newcommand{\rch}       {R_{\rm ch}}
\newcommand{\rct}       {{\tilde R_c}}

\newcommand{\tch}       {T_{\rm ch}}

\newcommand{\frat}      {F(25~\micron)/F(60~\micron)}

\begin{document}

%%%%%%%%%%%%%%
%Title Page %
%%%%%%%%%%%%%%

\lefthead{Chakrabarti et al.}
\righthead{}

\title{Feedback-Driven Evolution of the Far-Infrared Spectral Energy Distributions of Luminous and Ultraluminous Infrared Galaxies}

\author{
Sukanya Chakrabarti\altaffilmark{1,2},
T.J. Cox\altaffilmark{1}, Lars Hernquist\altaffilmark{1}, Philip F. Hopkins\altaffilmark{1}, Brant Robertson \altaffilmark{1}, Tiziana Di Matteo \altaffilmark{3}}

\altaffiltext{1}{
Harvard-Smithsonian Center for Astrophysics, 60 Garden Street, Cambridge, MA 02138 USA, schakrabarti@cfa.harvard.edu}

\altaffiltext{2}{
National Science Foundation Postdoctoral Fellow}

\altaffiltext{3}{
Carnegie Mellon University, Department of Physics, 5000 Forbes Ave., Pittsburgh, PA 15213}

\begin{abstract}

We calculate infrared spectral energy distributions (SEDs) from
simulations of major galaxy mergers and study the effect of AGN and
starburst driven feedback on the evolution of the SED as a function of
time.  We use a self-consistent three-dimensional radiative
equilibrium code to calculate the emergent SEDs and to make images.
To facilitate a simple description of our findings, we describe our
results in reference to an approximate analytic solution for the
far-IR SED.  We focus mainly on the luminous infrared galaxy (LIRG)
and ultraluminous infrared galaxy (ULIRG) phases of evolution.  We
contrast the SEDs of simulations performed with AGN feedback to
simulations performed with starburst driven wind feedback.  We find
that the feedback processes critically determine the evolution of the
SED.  Changing the source of illumination (whether stellar or AGN) has
virtually no impact on the reprocessed far-infrared SED.  We find that
AGN feedback is particularly effective at dispersing gas and rapidly
injecting energy into the ISM.  The observational signature of such
powerful feedback is a warm SED.  In general, simulations performed with
starburst driven winds have colder spectra and reprocess more of their
emission into the infrared, resulting in higher infrared to bolometric
luminosities compared to (otherwise equivalent) simulations performed
with AGN feedback.  We depict our results in IRAS bands, as well 
as in Spitzer's MIPS bands, and in Herschel's PACS bands.

\end{abstract}


%%%%%%%
%Body%
%%%%%%%

\section{Introduction}

Since the IRAS satellite detected the unexpectedly high amounts of
infrared emission generated by large numbers of dusty, infrared-bright
galaxies now commonly dubbed as LIRGs and ULIRGs (luminous infrared
galaxies, LIRGs, $L_{8~\micron-1000~\micron} \geq 10^{11} L_{\odot}$
and ultraluminous infrared galaxies, ULIRGs,
$L_{8~\micron-1000~\micron} \geq 10^{12} L_{\odot}$) (Soifer et al.
1984, 1987), there has been ongoing debate about the nature of the
infrared emission from these systems.  These galaxies are often
heavily obscured in the optical and radiate most of their energy at
mid and far-infrared wavelengths.  As such, much of our observational
understanding of these dusty galaxies has derived from analyses of
their infrared spectral energy distributions (SEDs).

A number of approaches have been adopted to study the infrared SEDs of LIRGs and ULIRGs.  
One common prescription has been to use infrared photometric templates (de Grijp et al. 1985, Xu et al. 2001, Egami et al. 2004, Lutz et al. 2005, Verma et al. 2005).  A key component of these templates is the now commonly used warm-cold IRAS classification developed by de Grijp et al. (1985), wherein sources with warm ($F(25~\micron)/F(60~\micron) \ga 0.2$) colors have been used to identify AGN.  It was not immediately clear from these templates what physical cause is responsible for this classification or how to interpret these templates within the context of self-consistent radiative transfer solutions.  The infrared emission for static density distributions has been
calculated self-consistently (Efstathiou \& Rowan-Robinson 1995, Silva et al. 1998) assuming that axisymmetric structures approximately represent the structure of galaxies.  More recently, approximate semi-analytical models of the evolution of galaxies, in particular the starburst component, have been included, and axisymmetric radiative transfer calculations (Granato et al. 2000, Efstathiou et al. 2000) and templates derived from such solutions (Farrah et al. 2002, Farrah et al. 2003) have been presented.  

However, observations paint a compelling picture of these galaxies
having a complex dynamical history, mediated by mergers and periods of
strong gas inflow and outflow (Soifer et al. 2000, Goldader et
al. 2002, Downes \& Solomon 1998, Scoville et al. 1998, Soifer et
al. 1999, Scoville et al. 2000, Surace et al. 2000, Surace \& Sanders
1999, Rupke et al. 2005a).  This scenario is supported by simulations
demonstrating that tidal interactions during a major merger cause gas
inflows by gravitational torques (e.g. Barnes \& Hernquist 1991,
1996), leading to nuclear starbursts (e.g. Mihos \& Hernquist 1994,
1996).  The early observations of the relics of mergers prompted
Sanders et al. (1988) to suggest a phenomenological model - where
ULIRGs would eventually become optically visible quasars once their
obscuring layers of dust were shed by supernova explosions, stellar
winds, and radiation pressure.  By and large, the common wisdom
regarding the warm-cold classification seems to derive from the idea
that the strength of the intrinsic radiation field from the AGN is
responsible for the higher frequency emission (Farrah et al. 2005,
Verma et al. 2005), and that other sources of radiation, like that
from a starburst, cannot be.

What is lacking so far is a self-consistent radiative transfer
calculation of the infrared emission of merging galaxies that
recognizes the wealth of observations that catalog their dynamically
active history.  Recent merger simulations incorporating
feedback from central black holes have captured in detail their highly
dynamical evolution (Springel et al. 2005a,b) and the relation between
mergers and quasar populations in optical and x-ray observations
(Hopkins et al. 2006b,c).  Moreover, simulations like these have been
successful at reproducing massive galaxies at high redshifts (Nagamine
et al. 2005a,b; Finlator et al. 2006; Night et al. 2006).
Recently, Cox et al. (2006b) have performed a
number of simulations with starburst driven wind feedback to analyze
the effects of this process during mergers.  These
simulations along with the simulations performed with AGN feedback
allow us to calculate the effects of AGN and starburst feedback on 
the time evolution of the infrared SED.

Our approach here in calculating the infrared SED improves on previous
work in two important respects: 1) we use information from simulations
that follow the dynamical evolution of merging galaxies without
using semi-analytic prescriptions as in prior work, and 2) we
have solved for the dust thermal emission self-consistently by using a
3-d radiative equilibrium code.  Our calculations here
are the $\it{first~self-consistent}$, i.e., wherein the dust temperature 
is calculated on the basis of radiative equilibrium, calculations 
of the infrared emission of dusty galaxies which utilize  
dynamical information from SPH merger simulations.  To facilitate a simple and intuitive
description of our results, we use an analytic solution for the far-IR
SED (Chakrabarti \& McKee 2005, henceforth CM05) to approximately
describe how the shapes of the emergent SEDs depend on the intrinsic
source parameters, namely the luminosity to gas mass ratio ($L/M$) and
the surface gas density ($\Sigma$).  Our goal in this preliminary
study is to broadly explore the large-scale differences between AGN
and starburst driven feedback on the SED.  We focus our discussion
here on the differences in the $\frat$ colors for these two kinds of
feedback.  We also depict our results in the Spitzer Space Telescope's MIPS and Herschel
Space Observatory's PACS bands.  We do not attempt to give a prescription to determine the
stage of evolution during the merger from the infrared colors.  We
investigate this problem, i.e., reverse engineering the stage of
evolution from multiwavelength colors, in a future paper, where we
calculate the entire spectrum, from X-rays to the infrared, for a
large set of merger simulations.

In \S 2, we review the methodology of the merger simulations and
describe how to translate the Smooth Particle Hydrodynamics (SPH)
information to the spatial grid that we use to do the radiative
transfer calculations.  We use a new prescription for the cold gas,
which is motivated by turbulent, rather than purely thermal support
for molecular clouds.  In \S 3, we review our methodology for the
radiative transfer calculations and the dust model we have adopted.
\S 4 is dedicated to an enumeration of our results.  In \S 5, we
present some qualitative arguments to describe the effect of feedback
on the gas distribution in the galaxy. We also point out the current
limitations of our treatment owing to the finite resolution of the
simulations, and preliminary analysis of radiative transfer
effects. \S 6 gives our main conclusions.

\section{Merger Simulations}

We employ a new version of the parallel TreeSPH code GADGET-2
(Springel 2005), which uses an entropy-conserving formulation of
smoothed particle hydrodynamics (Springel \& Hernquist 2002), and
includes a sub-resolution, multiphase model of the dense interstellar
medium (ISM) to describe star formation (Springel \& Hernquist 2003,
henceforth SH03).  The multiphase gas is pressurized by feedback from
supernovae, allowing us to stably evolve even pure gas disks (see,
e.g. Robertson et al. 2004). Black holes are represented by ``sink''
particles that accrete gas, with an accretion rate estimated using a
Bondi-Hoyle-Lyttleton parameterization, with an upper limit equal to
the Eddington rate (Springel et al. 2005b). The bolometric luminosity
of the black hole is then $\L_{\rm bol}=\epsilon_{\rm r}\dot{M}c^{2}$,
where $\epsilon_{\rm r}=0.1$ is the radiative efficiency.  We further
allow a small fraction ($\sim 5\%$) of $\L_{\rm bol}$ to couple dynamically
to the gas as thermal energy. This fraction is a free parameter,
determined in Di Matteo et al. (2005) by matching the $M_{\rm
BH}$-$\sigma$ relation.  We do not attempt to resolve the gas
distribution immediately around the black hole, but instead assume
that the time-averaged accretion can be estimated from the gas on the
scale of our spatial resolution, which is a few tens of parsecs in
the best cases.

We generate two stable, isolated disk galaxies, each with an extended
dark matter halo having a Hernquist (1990) profile and an exponential
disk. Our simulation is one of the series described in detail in Cox
et al. (2006a), with virial velocity $V_{\rm vir}=160\,{\rm
km\,s^{-1}}$, a fiducial choice with a rotation curve and mass similar
to the Milky Way.  We begin our simulation with 40 \% (by mass) gas
disks, which may correspond to local ULIRGs.  The galaxies are then
set to collide in a relative inclination of 60 degrees with zero
orbital energy and a pericenter separation of $7.1\,{\rm kpc}$.  We
study here the evolution of the galaxy in the main outflow phase,
close to the final merger of the two black holes.  The starburst
simulations we analyze are in all respects identical to the simulation
performed with AGN feedback (which we denote ``AGN'') except that the
feedback owes solely to starburst driven winds.  These simulations are
described in Cox et al. (2006b).  In these simulations,
starburst driven feedback is prescribed in terms of two independent
parameters, $v_{\rm W}$, the wind velocity, and the mass loading
efficiency of the wind $\eta$ (SH03).  We consider here three
starburst simulations which cover a range of parameters: SB1
($\eta=0.005,v_{\rm W}=837 ~\rm km/s$) has a very low mass loading
efficiency, SB9 ($\eta=0.05,v_{\rm W}=837 ~\rm km/s$), which has
moderate mass loading efficiency, and SB10 ($\eta=0.5,v_{\rm W}=837
~\rm km/s$), which has high mass loading efficiency and is nearly so
high that further increase would cause too much gas would be expelled
to produce a significant burst (as shown in Figure 10 of Cox et al. [2006b] -
which depicts the $\eta ~\rm vs ~v_{\rm W}$ two-dimensional parameter space; 
specifically, it demaracates the $\eta \ga 1$ region of parameter
space, where too much gas is removed for there
to be a significant burst, for any range of wind velocity).  
We are motivated to focus our discussion on $v_{\rm W} \sim 800 ~\rm km/s$ 
as the fiducial wind velocity, as the median
wind velocities observed in a large sample of LIRGs and ULIRGs by
Rupke (et al. 2005a) found similar median velocities.  
It is important to note that for our analysis here, we
consider either simulations with only AGN feedback (the ``AGN''
simulation) or simulations with only starburst driven feedback (the SB
series).

\subsection{Specification of the Interstellar Medium}

Owing to the limited resolution of current numerical simulations, the
interstellar medium (ISM) must be tracked in a volume-averaged manner.
In particular, the simulations discussed here adopt the multiphase
model of SH03, in which individual SPH particles represent a region of
the ISM that contains cold clouds embedded in a diffuse hot medium.
Because the SPH calculation uses only the volume-averaged pressure,
temperature, and density to evolve the hydrodynamics, this model does
not provide specific information regarding the cold clumps. Since
these cold clumps harbor the dust that produces the infrared emission,
we must adopt a model which determines their number, density, and
size, and location.  (Our approach is similar to that employed
by Narayanan et al. [2006] in their study of the evolution of
the molecular gas in mergers.)

As the first step to specify the properties of the cold, dense gas, we
assume that each SPH particle contains one dense molecular cloud at its
center.  Because we know the temperature and density of the volume-average
SPH particle, assuming a temperature (say, $10^{3} \rm ~K$, as was assumed in
SH03) for this cold cloud and that it exists in
pressure equilibrium with the hot phase we readily attain its density.
However, observations of molecular clouds indicate that their pressure is
dominated by turbulent motions, rather than thermal energy, and thus
neglecting this feature yields very dense clouds, with small volume
filling factors.  In the work presented here, we take the turbulent
pressure to be $\sim 100$ times that of the thermal pressure.  (We took
this ratio to be 100 in the simulations presented here - we have also
varied this ratio by a factor of several on either side to find no qualitative
differences).  With this
assumption, the density is much lower than without the turbulent support
and its volume filling factor is greatly increased.  

This prescription for turbulent support is consistent with observations and models 
of GMCs.  In a recent review on GMCs, Blitz et al. (2006) note that the size-line width 
relation observed in our galaxy (Solomon et al. 1987) owes to the turbulent nature of the gas inside GMCs and agrees roughly with measurements of extragalactic GMCs.  In a detailed study 
of the starburst galaxy M64, Rosolowsky \& Blitz (2005) find that the GMCs are
virialized ($GM \sim 5\sigma^{2} R$), but that there are some key differences
in the line-width size relation from that found by Solomon et al.. (1987) for the 
Milky Way.  The average linewidth they observe ($\Delta V=28~\rm km/s$
by averaging the 25 entries for their resolved clouds in their Table 3) 
gives a similar ratio of turbulent to thermal pressure for $T=30~\rm K$ gas.  
Solomon et al.'s (1997) study of the GMCs in ULIRGs found larger linewidths
still, but this is clearly an upper bound since the individual clouds are
not resolved, and large scale motions, rather than internal motions 
will have contributed to the measurements of the line widths.  Our adopted
ratio of turbulent to thermal pressure is close to that for the interpretative
``mist model'' for GMCs proposed by Solomon et al. (1987).  Therefore, although the 
details of the variations in the size-line width relations in extragalactic GMCs are
not yet observationally clear, it is well-established that they are virialized and
supersonically turbulent.  As such, the inclusion of turbulent pressure in modeling
the sizes of the cold clumps is observationally well justified.  Our specification for the ISM
is also in agreement with recent theoretical work - Krumholz \& McKee (2005)  
derived a simple prediction for the star formation rate that agrees with 
the Kennicutt-Schmidt law for a diverse range of galaxies, from spiral galaxies to ULIRGs, 
by requiring that star formation occurs in virialized molecular clouds 
that are supersonically turbulent.    

\section{Methodology: Radiative Transfer}

We use a self-consistent three-dimensional Monte Carlo radiative equilibrium code (Whitney et al. 2003), generalized to include distributed sources of radiation, to calculate the emergent SEDs and images from the merger simulations as a function of evolutionary state.  The details of this generalization and its effects on the SED are discussed in a separate paper (Chakrabarti \& Whitney 2007, in preparation).  We review some the main points
here briefly.  This code incorporates the Monte Carlo radiative equilibrium routine developed by Bjorkman \& Wood (2001) to solve for the equilibrium temperature of the dust grains.  Individual photons are tracked directly in the Monte Carlo scheme.  When a photon is absorbed by a grid cell, it may raise the temperature of that grid cell so that it emits infrared radiation; the emissivity of this grid cell depends on the temperature.  The Bjorkman \& Wood (2001) algorithm corrects the temperature in a grid cell by sampling the new photon frequency from the difference of two emissivity functions - that from the previous temperature and that computed with the new temperature.  While this is not an explicitly iterative scheme, a sufficiently large number of photons must be absorbed by the grid cells so that they eventually relax to an equlibrium temperature.  Since the density within a grid cell is constant, one can easily integrate the optical depth through the grid to account for the attenuation of the photons.  This code is capable of treating scattering as well, which is particularly relevant for the shorter wavelengths.  Since we focus on the far-IR here, and the scattering efficiency varies as $1/\lambda^{4}$, we have neglected scattering in these calculations.  In a forthcoming paper on panchromatic SEDs (which span the range from optical to millimeter wavelengths) of submillimeter galaxies (Chakrabarti et al 2006b), we include scattering along with the absorption and re-emission processes to calculate the emergent SEDs.  Whitney et al. (2003) carried out convergence studies to compare their results to a set of benchmark calculations developed for spherically symmetric codes.  Chakrabarti \& Whitney (2007, in prep) have carried out both photon and grid resolution convergence studies to verify that photon numbers of order $50 \times 10^{6}$ yield converged results from mm to near-IR wavelengths for the three-dimensional grids we have used here; the grid resolution was varied by almost an order of magnitude to find reasonably converged results.  For fine grid structures, it is necessary to increase the photon number proportionately to ensure that each grid cell receives enough photons to reach an equlibrium temperature (i.e., the photon and grid resolution studies are not independent).  Off-nuclear sources of radiation are allocated photons in proportion to the fraction of bolometric luminosity they contribute.  The radiative equilibrium temperature calculation needs to be referenced to the total bolometric luminosity of all of the sources of radiation.  For large three-dimensional grids, it is crucial to allocate tens of millions of photons for the grid cells to reach an equilibrium temperature; otherwise, the emissivity of the dust may be artificially lowered.  This code incorporates a ``peeling-off'' algorithm (Yusef-Zadeh, Morris \& White 1984; Wood \& Reynolds 1999) to calculate the images, which are then convolved with broadband filter functions.     

The merger simulations, as discussed previously, directly track the mean SPH gas density.  We take the breakdown in the cold gas phase to represent the dust spatial distribution, i.e., we assume that the dust and gas are coupled.  We have used a logarithmic grid, with $\Delta r/r \sim \Delta \theta/\theta \sim \Delta \phi/\phi \sim 0.02$.  The grid resolution is set to resolve the average size of the cold gas clumps, particularly on scales of the Rosseland photosphere, as we discuss below.  The Rosseland photosphere can be understood as the effective $\tau=1$ surface at the peak frequency of the SED.  The outer radius is set to 10 kpc, as densities beyond several kpc are very low ($\rho < 10^{-25} ~\rm g/cm^{3}$) and as such do not contribute to the emitted far-IR spectrum.   

We model the intrinsic AGN continuum spectrum following Marconi et
al. (2004), which is based on optical through hard X-ray observations
(Elvis et al. 1994, George et al. 1998, Perola et al. 2002, Telfer et
al. 2002, Ueda et al. 2003, Vignali et al. 2003), with a reflection
component generated by the PEXRAV model (Magdziarz \& Zdziarski 1995).
Since we focus on the reprocessed far-IR SED here, the details of the
intrinsic AGN spectrum as well as the intrinsic starburst spectrum
are, for our purposes, not of much consequence.  As discussed in CM05,
if the source of radiation is obscured, as is the case here, the
reprocessed spectrum is approximately independent of the intrinsic
stellar (or AGN) spectrum.  The stars composing the starburst in our
calculations have an effective stellar temperature of 15,000 K - this
is a fair approximation given that young stellar populations are
expected to dominate the luminosity in the starburst (Bruzual \&
Charlot 1993).  While we track photons from all of the stellar
particles in the simulation which are distributed throughout the
grid, the starburst in the simulations, which
produces most of the luminosity from the stars, is generally compact - of 
order a couple of hundred pc.  (At late times, the starburst is somewhat more diffuse). 
The compact sizes of the starbursts we find in the simulations are consistent with 
observations - high spatial resolution observations of the nearest ULIRGs by
Soifer et al. (2000) shows that most of the mid-IR emission originates
on scales of $\sim 100$ pc.  If the mid-IR emission is powered by a starburst,
this suggests that the starburst is compact, and on scales of $\sim 100$ pc.
We have numerically verified that little change ($\sim
10 \% $) results in the far-IR SED even when the stellar temperature
(spectrum) is changed by a factor of three.  To understand the near-IR and
polycyclic aromatic hydrocarbon (PAH) spectrum, it is necessary to carry out
stellar population synthesis modeling, as these parts of the emitted
spectrum do depend on the intrinsic radiation field.

As is now well known, very small grains are subject to temperature
fluctuations, so that they may not be in equilibrium with the thermal
radiation field, if the amount of energy striking the grain is larger
than its specific heat capacity (Purcell 1976).  This leads to a
probability distribution function for the temperature (rather than one
temperature) for grains of size less than $\sim 0.01~\micron$
(Guhathakurta \& Draine 1989, Siebenmorgen et al. 1992, Manske \&
Henning 1998).  PAH heating has been
treated in a way similar to the heating of very small dust grains
(Manske \& Henning 1998).  Observationally, we do find strong PAH
emission for $\lambda < 8 ~\micron$, particularly in LIRGs (Rigopoulou
et al. 1999, Brandl et al. 2004, Spoon et al. 2004, Yan et al. 2005),
but little discernible PAH emission at longer wavelengths.  Moreover,
given that PAH emission is essentially a quantum process, i.e., it
owes to the absorption of single photons and their subsequent effect,
PAH emission arises only from very small grains.  Such small grains
and the details of their temperature fluctuations are unlikely to
affect the far-IR spectrum.  Therefore, in this paper, we use a
simplified representation of PAH emission instead of calculating it
self-consistently - if there is a UV photon of sufficient magnitude
to excite PAH emission, then we add on an observed PAH template
(adopted from Brandl et al. 2004) to the SED.  We investigate in
detail the near-IR spectrum and PAH emission in a future paper, and
consider the effects of scattering of photons off dust grains, the AGN
spectrum, using stellar synthesis models to compute the stellar
spectra.

We use the Weingartner \& Draine (2001) (henceforth WD01) $R_{V}=5.5$
dust opacity model.  The mass fraction of dust is equal to $1/105.1$
for solar abundances (WD01).  The long wavelength ($\lambda > 30 ~
\micron$) part of WD01's grain model has the same slope ($\beta=2$) as
the Draine \& Lee (1984) dust model.  WD01's models have been shown to
reproduce the observed extinction curves for the Milky Way, as well as
regions of low metallicity, such as the LMC and the SMC.  The opacity
normalization per gram of dust, $\kappa_{\lambda_{0}}$,is equal to
$0.27 \delta$ for $\lambda_{0}=100~\micron$, where $\delta$ is the
dust-to-gas ratio relative to solar, which we take to be unity.  Dunne
\& Eales (2001) and Klaas et al. (2001) found that the dust-to-gas
ratio for a large sample of ULIRGs is comparable to Milky Way values,
when they fit two-temperature blackbodies to the far-IR SEDs.
Previous work, based on fitting single temperature blackbodies had
found slightly lower dust-to-gas ratios (Dunne et al. 2000) by a
factor of two.  The basic reason for this is that using a single
temperature for the dust forces the fit to shallower opacity slopes
($\beta\sim 1$) to match the long wavelength part of the SED.  Because
the derived dust mass for modified blackbody models is proportional
$(3+\beta)$, this results in somewhat lower dust masses (Chakrabarti
\& McKee, in preparation).

CM05 developed the relations between the source parameters, $L/M$ and
$\Sigma\equiv M/\pi R^{2}$ (proportional to the column), and the SED
variables, $\rct$ and $\tch$ that govern the shape of the SED for a
spherically symmetric geometry.  The SED variable $\rct$ is the ratio
of the outer radius (beyond which the density drops sharply) to an
effective Rosseland photosphere ($\rch$); $\tch$ is the temperature at
the effective Rosseland photosphere.  Dust envelopes characterized by high
$L/M$ values have higher photospheric temperatures and peak at shorter 
wavelengths; low $\Sigma$ envelopes are more extended (larger $\rct$), 
the SEDs of which do not fall as sharply, shortwards of the peak wavelength, 
as high $\Sigma$ sources.  Both of these effects increase the ``warmness'' of
the SEDs, by shifting the peak to shorter wavelengths (increasing $L/M$) and
by lowering the effective attenuation of high frequency photons (decreasing $\Sigma$). 
In Figure 5, we depict the infrared
luminosity (from integrating the SED from $8~\micron$ to
$1000~\micron$), the bolometric luminosity, and $\Sigma$ (taking
the mass interior to 10 kpc) values 
for the cases we discuss here.  For
typical $L/M$ and $\Sigma$ values during the simulation, this gives a
Rosseland photosphere of order 100 pc, somewhat larger than the
hydrodynamic resolution scale of the simulations employed here.  This
is an approximate estimate of the Rosseland photosphere, since CM05's
formalism did not consider the effects of clumping, distributed
sources, or nonspherical geometry.

\section{Results:  SEDs \& Images of Simulations with Starburst and AGN Feedback}

We now discuss and contrast our results for the simulation with AGN
feedback and the simulations with starburst feedback, for times close
to the main outflow phase.  The images in the MIPS bands of the
Spitzer Space Telescope (at $24~\micron,70~\micron,160~\micron$) are shown in Figures 1-4 for three times
during the simulation - $t=1.165 ~\rm h^{-1}Gyr,t=1.205~\rm h^{-1}Gyr,t=1.33~\rm h^{-1}Gyr$,
which correspond to times before the main outflow phase, during the outflow phase,
and close to the end of the outflow phase respectively.  The time
is given in terms of $h^{-1} \rm Gyr$, with $h=0.7$.
The green traces the MIPS $160~\micron$ emission, the red the $70~\micron$, and the purple the
$24~\micron$.  We have adopted a distance of 77 Mpc in all of these
figures and an effective angular resolution of $5''$.  As such, these figures 
should be interpreted as visual aids and not realistic synthetic images.  
Figure 1a shows the image of the obscured AGN phase.  Figure
1b shows the image at a time 40 $\rm Myr$ after Figure 1a, when the
outflow has pierced through the obscuring envelope and cleared out
lines of sight.  Figure 1c shows the images another 125 $\rm Myr$
later, when the outflow has cleared out most of the lines of sight.
For comparison, the observable quasar lightcurves of this and other
similar simulations are shown in detail in Hopkins et
al. (2005a,2006a) and trace a similar evolution with time and outflow
phase from a heavily obscured to optically visible quasar.  Figures
2, 3, and 4 depict the simulations with starburst driven feedback.
Figure 5 shows the detailed time evolution of the ratio of 
the infrared to bolometric luminosity, along with the 
total infrared luminosity, and the time variation of the gas
surface density.  Figure 5a shows the time dependence of the ratio of the infrared luminosity (from $8~\micron-1000~\micron$) to the total bolometric luminosity.  The simulations performed with starburst feedback generally have higher $L_{\rm IR}/L_{\rm bol}$ since the optical extinctions are higher and cause more of the emission to be reprocessed into the infrared.  
An important point to note from Figures 5b and 5c is that
the simulation with AGN feedback (solid line) has lower
columns ($\Sigma$) (Figure 5c) and higher luminosities (Figure 5b) compared to the SB series.
In particular, the AGN simulation loses mass quickly during
its luminous phases ($t=1.205 h^{-1}~\rm Gyr$), while starburst
feedback disperses gas more gradually (Figure 5c).  As there is often
much debate as to how much the black hole contributes to the total 
bolometric luminosity, we also depict in Figure 6 
the ratio of the luminosity of the black hole relative to the total bolometric 
luminosity.  As this plot shows, the black hole dominates the contribution to the
bolometric luminosity only for a short period of time during the active phase; this trend 
is generally observed across a large number of simulations.

The images in Figures 1-4 show that the contribution function or characteristic
emission radius varies with wavelength.  CM05 calculated the
contribution function for a spherically symmetric, homogeneous
envelope and showed that the longer wavelengths emanate from
the outer cool parts of the envelope, while the shorter wavelengths
come from deeper inside the envelope, modulated by the competing 
effects of optical depth and the temperature gradient.  Figures 1-4 illustrate
this point - the $24~\micron$ emission, shown in purple, comes from 
the inner regions ($\la \rm 1 kpc$) while the $160~\micron$ emission 
comes primarily from the outer parts of the dust envelope. The images 
also show the disruptive effects of the outflow on the
large scale structure of the galaxy (scales of order and larger than a
kpc).  The spatial distribution of the long wavelength
emission in the $160~\micron$ band, which traces
the large scale structure, changes considerably.  However, owing to
our current lack of resolution, the details of the shorter wavelength
MIPS band ($24~\micron$, shown in purple), which comes from deeper in
the dust envelope, is not traced as well.  The ``AGN'' simulation has a higher
bolometric luminosity than the SB series - so the images in all the
MIPS bands are correspondingly brighter.  

The SEDs of the simulation
with AGN feedback and the simulations 
starburst driven feedback (the SB series) are shown in Figures 7.
We have added on a observed PAH template when the conditions described
in \S 3 are satisfied.  It is important to note that this is not a
self-consistent treatment of PAH effects, and can at best be
considered as a lower bound on the actual PAH emission.  It does
indicate however, that PAH emission is most likely to be found in
LIRGs rather than ULIRGs since the extremely high levels of dust
emission in ULIRGs do not allow for the PAH spectral features to be
visible above the continuum.  The key point to note from Figure 6
is that the SEDs of ``AGN'' are broader in the most luminous phase
(t=1.1205 Gyr, the solid line) and do not fall as sharply on the Wien
end as the SB series - this may be explained in terms of the relative
difference in gas surface density, which for the ``AGN'' simulations
is also coincident with the increase in $L/M$.
Another point to take note of from these figures is that there is a
significant decrease in luminosity (of a factor of 10) from $t=1.205 h^{-1} \rm
Gyr$ and $t=1.330 h^{-1} \rm Gyr$ for the AGN run, with a similar trend for the SB
series.

Figure 8 shows the evolution of the $F(25~\micron)/F(60~\micron)$
colors as a function of time.  SEDs are classified as ``warm'' for
$F(25~\micron)/F(60~\micron) \ga 0.2$.  The AGN simulation is generally
warmer than the simulations with starburst feedback, and particularly
so in the most luminous phase ($t=1.2 h^{-1} ~\rm Gyr$) (for a qualitative
discussion of this, see \S 5).  The warm SED correlates with the
strong outflow phase, which suggests that feedback effects are
responsible for the increase in the high frequency emission.  
Figure 8 also shows that there is a general trend for the SEDs to
become warmer as the mass loading efficiency of the starburst winds
increases - this trend also suggests that the warm-ness of the
spectrum owes to feedback.  At late times, all the simulations lose
cold gas mass, either owing to gas dispersal by the outflow or to
new star formation - this eventually leads to the colors becoming
colder since there is not as much gas left to heat up to high
temperatures.  One important point to take note of here is that even if the source is unresolved, i.e., even if we could not construct images of the source at all wavelengths (i.e., those where the spatial resolution of the instrument does not allow us to) to see the variation of the surface brightness, the variations in $\frat$ give a direct clue as to the dynamical evolution of the system.  The magnitude of the $\frat$ colors will depend also on the clumpiness internal to SPH particles, which cannot be treated here owing to the finite resolution of the simulations.  The exact phase of evolution when the SEDs will become
cold will depend on the mass of the galaxies - as more available gas
will fuel the warm phase for a longer time.  Since we consider a small
subset of simulations here, we do not attempt to quantify the relative
amount of time that is spent in the warm phase.  We address these
questions statistically in a future paper by analyzing a large number
of simulations.

In Figure 9a, we show the observed colors in the MIPS band for local 
galaxies, and in the observed frame for galaxies at $z=1$ in Figure 9b.  
For observed frame colors of $z=1$ systems, ``warm'' corresponds to
$F_{70~\micron}/F_{160~\micron} \ga 0.6$ - in the rest frame,
these bands are now probe a slightly different region of 
the spectrum. We also depict the cold-warm 
trend in the PACS bands of Herschel in Figures
10.  In $F_{75~\micron}/F_{110~\micron}$ for systems at $z=1$, ``warm''
would correspond to $F_{75~\micron}/F_{110~\micron} \ga 0.6$ and 
for systems at $z=2$, $F_{110~\micron}/F_{170~\micron} \ga 0.6$, 
in the observed frame.  Future observations by these missions
may help to discover energetically active AGN by searching for
``warm'' systems in these bands.

\section{Discussion}

We can understand the differences in the emergent SED for AGN and
starburst driven feedback simulations in a simple way by considering
the relative energy injection per time (the power) for these two kinds
of feedback mechanisms.  The energy injection by supernovae is given
by: $E_{\rm SN}=(E_{0}/M_{0})M_{\rm SN}$ where $E_{0}$ and $M_{0}$ are
$10^{51}~\rm ergs$ and $8 M_{\odot}$ respectively.  Taking $M_{\rm
SN}=\beta M_{\star}$, where $\beta\sim 0.1$ (e.g., SH03), and taking
the supernovae energy to couple with efficiency of order unity
($\eta_{SN}=1$), we see that $E_{\rm SN}^{\rm
coupled}=(E_{0}/M_{0})\beta M_{\star}$.  The black hole energy couples
with efficiency $\eta_{BH}$ which is of order 0.1 
(Springel et al 2005).  The amount of energy then that couples
to the ISM from the black hole is given by:
$E_{BH}=\eta_{BH}\epsilon_{\rm r}M_{BH}c^{2}$, where $M_{BH}=0.001
(M_{\star}/f_{\rm SB})$ where $f_{\rm SB}$, the fraction of the stars formed 
in the starburst is of order 0.1 (Marconi \& Hunt 2003).  With the above values,
$E_{SN}^{\rm coupled}/E_{BH}^{\rm coupled}$ evaluates to order 0.1.  If the
coupling efficiency for supernovae is less than unity, as has been considered
by Recchi \& Matteucci (2002), it would be even more difficult to 
reconcile starburst driven feedback with a warm far-IR SED.
The key difference in how this energy affects
the medium owes to the timescales over which this energy is injected.
The timescale for supernovae feedback from the starburst is roughly
several hundred Myr, while the timescale for the black hole is the
Salpeter time, less than 40 Myr.  In essence then, the black hole
delivers more power to the ISM, and has a sharper effect on the
surrounding medium since energy is delivered rapidly enough to prevent
effective cooling.  This is graphically shown in Figure 11.

Our model of wind feedback is approximate - the parameters that we
have taken to describe starburst feedback, i.e., the wind velocity and
the mass loading efficiency may well depend on the microphysics of the
ISM so as to be a function of both time and spatial position.
However, it is worth emphasizing that the parameters that we use to
prescribe the feedback are broadly consistent with observations.  In a
comprehensive recent survey of winds in LIRGs and ULIRGs, Rupke et
al. (2005a,b) find that wind velocities are generally of order
$600~\rm km/s$ and the mass loading efficiency is of order 0.1.  The
mass loading efficiency is often taken to be unity in numerical
simulations (Kauffmann \& Charlot 1998, Aguirre et al. 2001a,b).
We have varied the mass loading efficiency (from 0.005-0.5) to find
significant changes result - in terms of the mass loading efficiency,
these observations indicate that the SB9 and SB10 models may be the
most realistic amongst the ones studied here from the SB series.

One of the main points in our presentation is that feedback
effects (either from starburst winds or AGN) may well be
responsible for warm SEDs.  This point is evinced by the the increase
in the warm-ness of the spectrum that we observe as we increase the
amount of feedback (through increasing the mass loading efficiency).  
The effective mass loading efficiency for AGN feedback is considerably
higher than for starburst feedback and is discussed in detail in Cox et al. (2006b).
Moreover, the trends that we see as a function of time are intuitive -
the spectrum goes from cold to warm as the outflows begin and the
column of obscuring dust and gas is lowered.  We have observed these
trends in other simulations as well.  These two points, taken
together, suggest that feedback effects are responsible for the
production of warm SEDs.

We have focused here on changes in the SED that are tied to dynamical
effects which are tracked well by the merger simulations, i.e., the
density structure on scales larger than the SPH smoothing length.  Small
scale structures, such as clumpiness of the ISM on scales of order 
tens of pc or smaller are not currently directly tracked in the simulations.
It is possible that higher resolution simulations will find
quantitatively different results, if they are able to better resolve
the clumpiness on small scales.  However, since clumpy structures are
generally coincident with the outflow phase, as studied in other
detailed numerical simulations of outflows (Stone et al. 1995), it is
unlikely that the trends as a function of time will be different.  In
other words, resolving the clumpiness of small scale structures may
lead to even higher frequency emission close to the outflow phase,
since the optical depth of a clumpy medium is lower than that of an
equivalent homogeneous medium (Natta \& Panagia 1984).  However, the
details of the effects of clumping on the spectrum are not yet clear -
although the optical depth of a clumpy medium is lower, this 
may be (partially) offset by the denser clumps trapping radiation more
effectively.  It is difficult to see how clumpiness of the ISM could
alter the trends as a function of time without resorting to some
ad-hoc (rather than dynamically motivated) changes in the small scale
structure of the gas.  However, since we have not tracked the
small-scale structures - we cannot present a definitive analysis of
the effects of clumping on the spectrum here.  Therefore, while we
find that warm SEDs are correlated with the outflow phase, we cannot
rule out other processes producing warm spectra.

Finally, the trends that we observe in the simulations agree with
observations.  There is a general tendency for higher luminosity
systems to be powered by AGN (Lutz et al. 1998, Tran et al. 2001,
Charmandaris et al. 2002).  However, the notion that AGN generate warm
SEDs because the intrinsic hardness of the radiation field produces
hot dust does not seem warranted.  Temperatures of order
100 K (and higher) are often inferred in protostellar regions, which
are powered by single stars or a stellar cluster (see e.g. Sridharan
et al. 2002, Mueller et al. 2002, de Buizer et al. 2005, among recent
references on massive protostellar regions) and have significant
amounts of high frequency emission.  Although the work by Rupke et
al. (2005a,b) suggests that mass loading efficiency of order 0.1 is
expected for local LIRGs and ULIRGs, Erb et al. (2006) deduce mass
loading efficiencies larger than unity, which if valid, could in
principle be enough for starbursts to produce warm SEDs, but which
may suppress star formation below the ULIRG threshold.

We also note here that Hopkins et al. (2005b,2006a) analyzed a large number of simulations to find similar trends in the quasar light curves, with sharper features for the more massive systems.  It will be useful to discern the phase of evolution from multiwavelength colors for
quantitative interpretation of observations within the context of a dynamical
model.  In a future paper, we give a prescription for estimating the phase of evolution by analyzing the multiwavelength spectra (from X-rays to infrared) of a large suite of simulations designed to be representative of the local ULIRG population.  We also extend this analysis to more gas rich and massive systems which may be representative of high redshift systems to quantify the statistics of sub-mm galaxies and bright X-ray and infrared sources (Chakrabarti et al 2006b).   

\section{Conclusion}

We have used a three-dimensional self-consistent radiative equilibrium code to compare the SEDs of simulations performed with only starburst feedback with simulations that include only the effects of AGN feedback.  The main points are:

1) We find that feedback, either from the AGN or starburst driven winds, is very likely the dynamical agent that is responsible for changing the shape of the far-IR SED, in particular, the $F_{\lambda}(25~\micron)/F_{\lambda}(60~\micron)$ colors.  We also depict this trend 
in Spitzer's MIPS bands and in Herschel's PACS bands.

2) We find that, simulations performed with AGN feedback have SEDs that are warmer, particularly in their most luminous phases, relative to equivalent starburst driven simulations.  The basic reason for this general difference is that the sharp energy injection from AGN over short timescales disperses gas efficiently resulting in lower columns of dust during the peak of luminosity.  

3) The fraction of infrared to bolometric luminosity is generally higher for starburst driven simulations.  For all other things being equal, starburst driven simulations have lower total bolometric luminosities compared to simulations with AGN feedback.

4) The presence of AGN or any source of illumination cannot be inferred from the far-IR SED alone if the source of illumination is heavily obscured such that the far-IR arises predominantly from the reprocessed emission of thermally heated dust grains.  The sources of illumination (namely the AGN and the compact starburst) in the merger simulations are generally sufficiently obscured in the ULIRG phase of evolution such that this condition is met.  This suggests that the warm, cold classification of the SEDs of LIRGs and ULIRGs is independent of the source of illumination.

5) Simulations performed with AGN feedback generally have lower $\Sigma$ (lower columns) and higher $L/M$ than otherwise equivalent simulations with starburst driven feedback, and as such have more flux on the Wien part of the spectrum.

We thank Christopher McKee for many helpful discussions, particularly on the interstellar medium, which have motivated our prescription for the ISM.  We thank Jennifer Hoffman and especially
Barbara Whitney for many helpful discussions on Monte Carlo radiative
transfer, in particular on the treatment of multiple sources.  We also thank Erik Rosolowsky for informative discussions on the structure of extragalactic GMCs.  We also thank an anonymous referee for a careful reading of the manuscript and helpful suggestions.  The research of SC has been 
supported by a NSF postdoctoral fellowship.  The calculations have been performed on the Institute for Theory and Computation cluster.  

\clearpage

\begin{figure}[h] \begin{center}
\centerline{\psfig{file=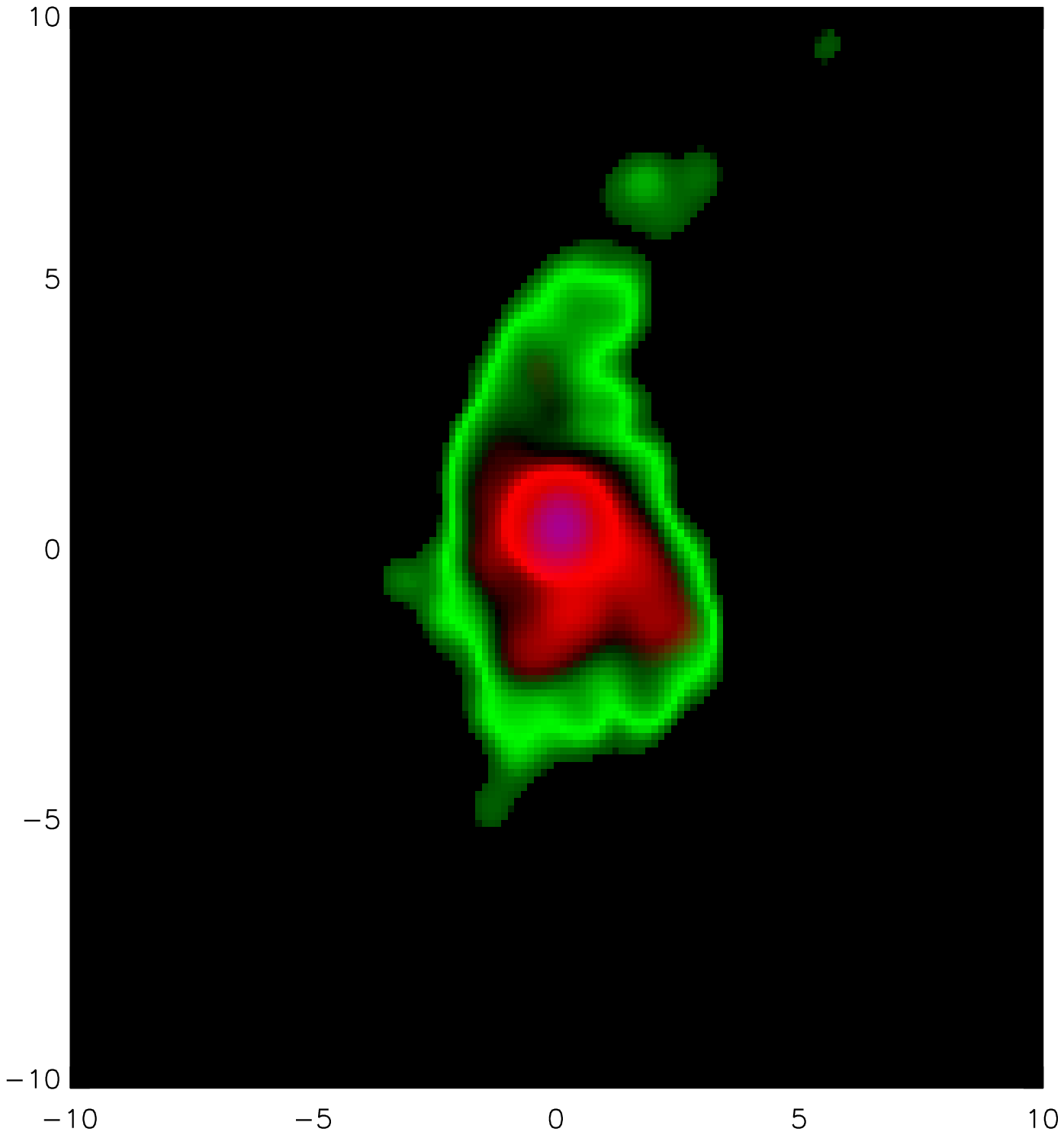,height=2.in,width=2.in}
\psfig{file=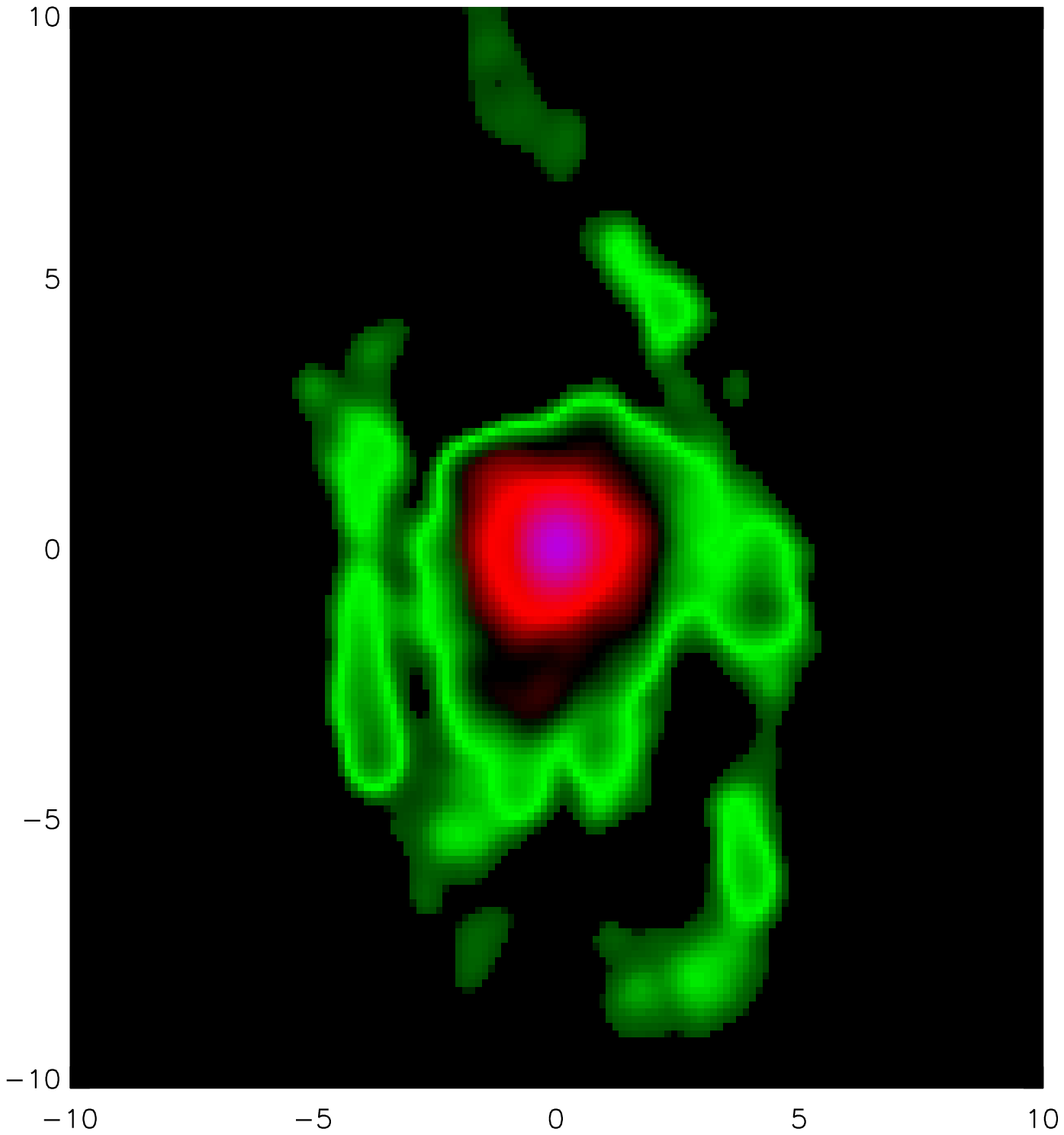,height=2.in,width=2.in}
{\psfig{file=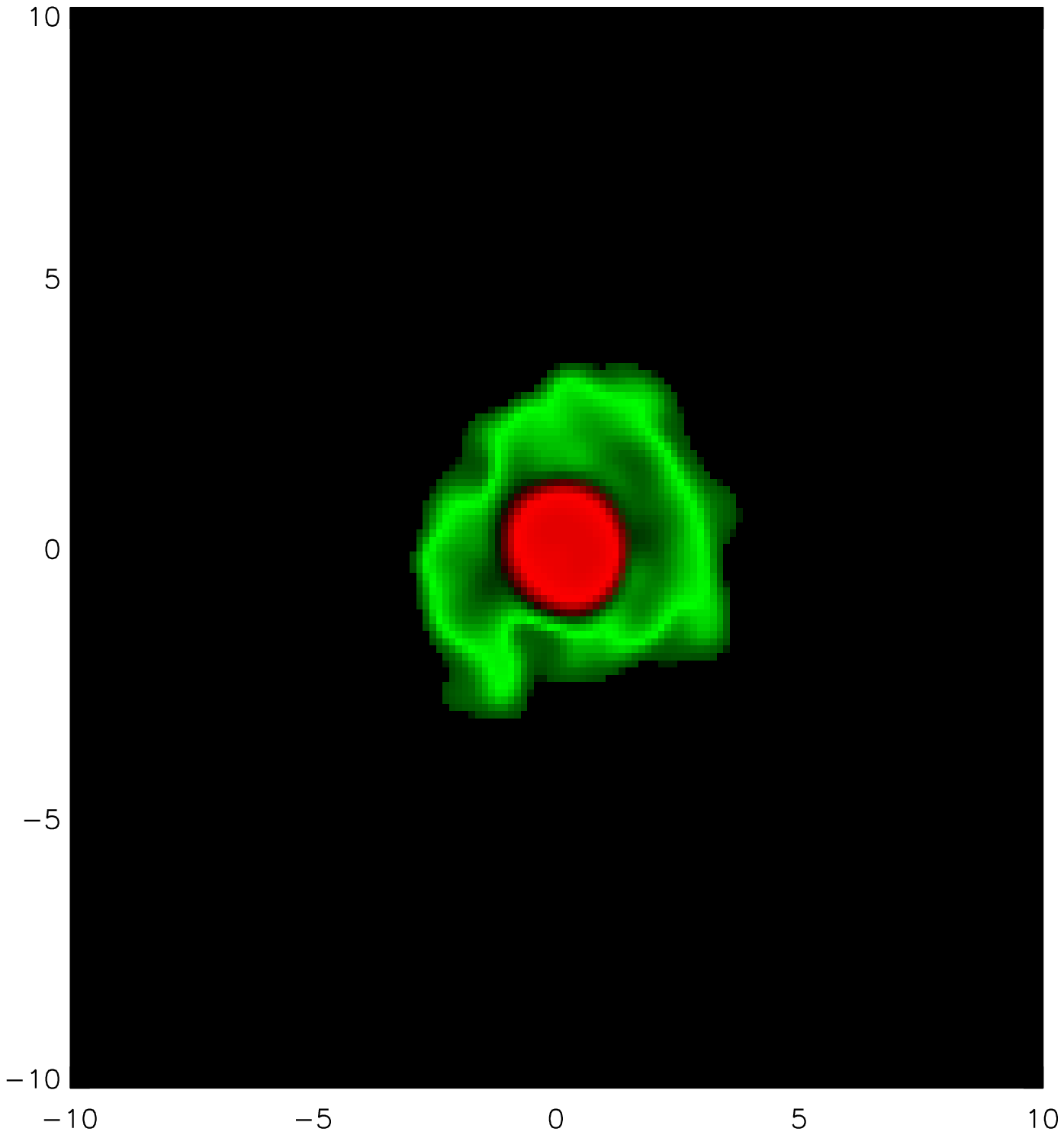,height=2.in,width=2.in}}}
\end{center}
\caption{(a) Obscured AGN phase ($t=1.165 h^{-1}\rm Gyr$), (b) Outflow starts to clear out lines of sight ($t=1.205 h^{-1}\rm Gyr$), i.e., the ``warm'' phase.  Note that the $24~\micron$ flux (shown in purple) peaks in this phase. (c) Close to end of outflow phase ($t=1.330 h^{-1}\rm Gyr$)}
\end{figure}

\begin{figure}[!hb] \begin{center}
\centerline{\psfig{file=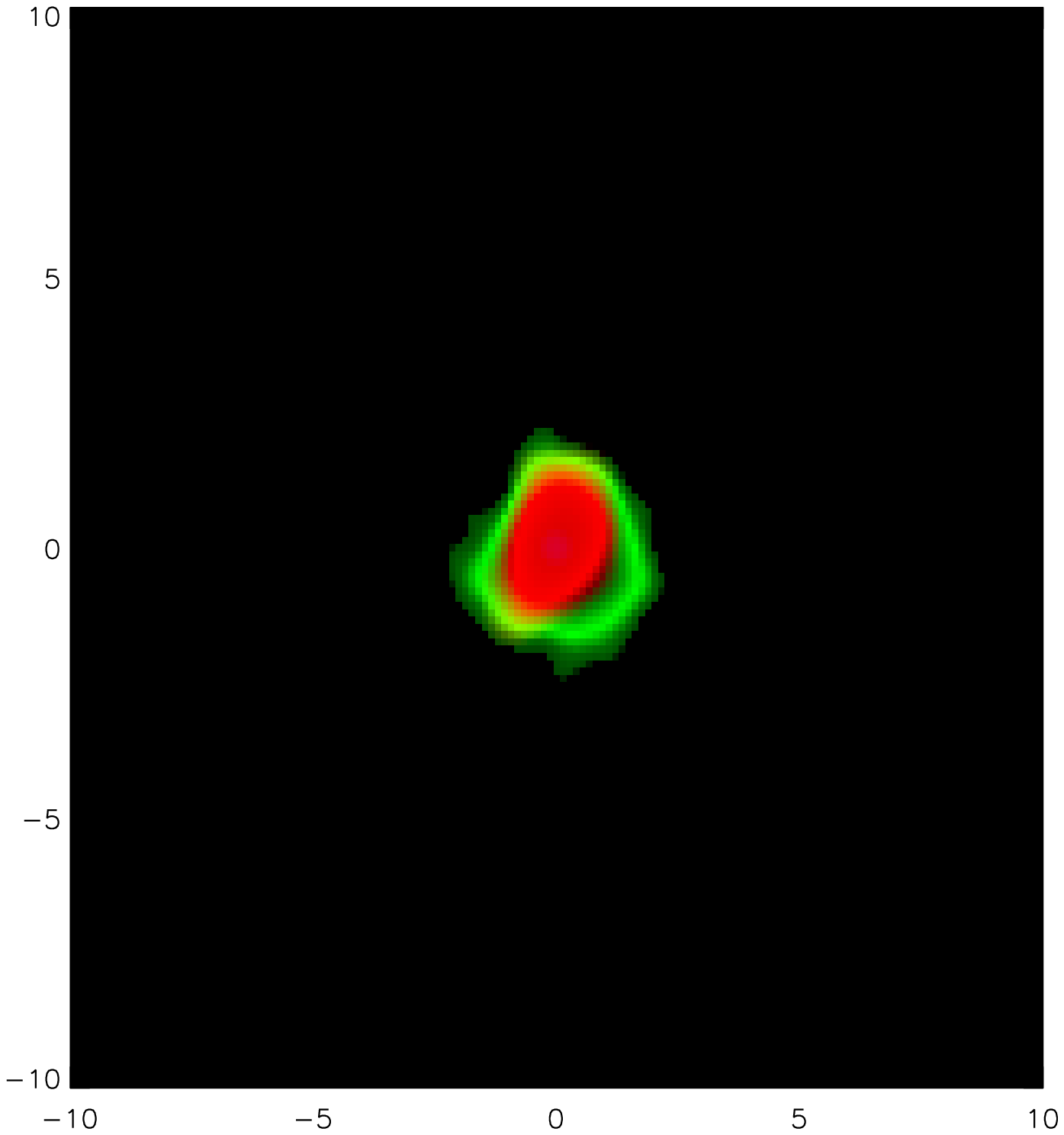,height=2.in,width=2.in}
\psfig{file=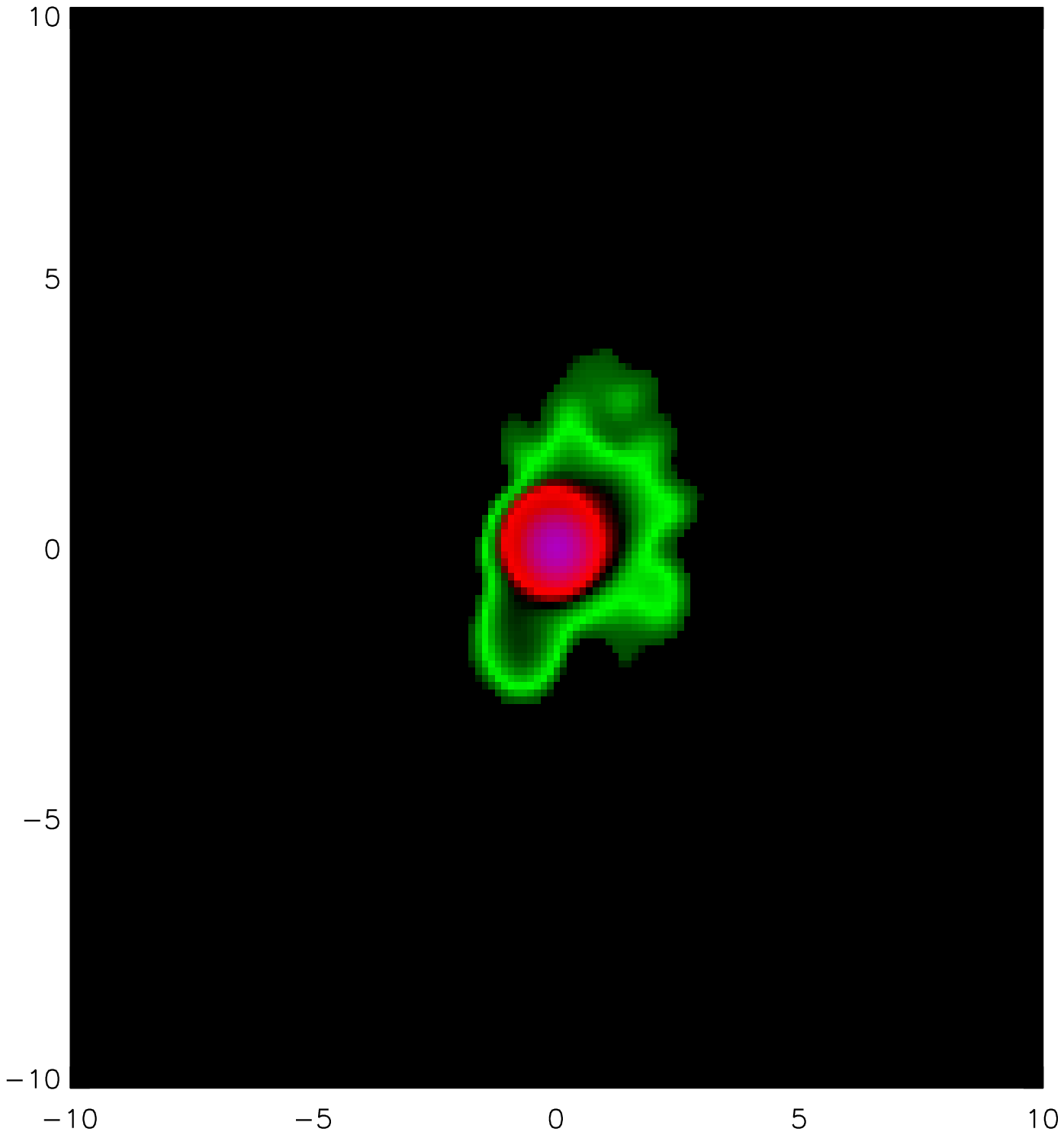,height=2.in,width=2.in}
{\psfig{file=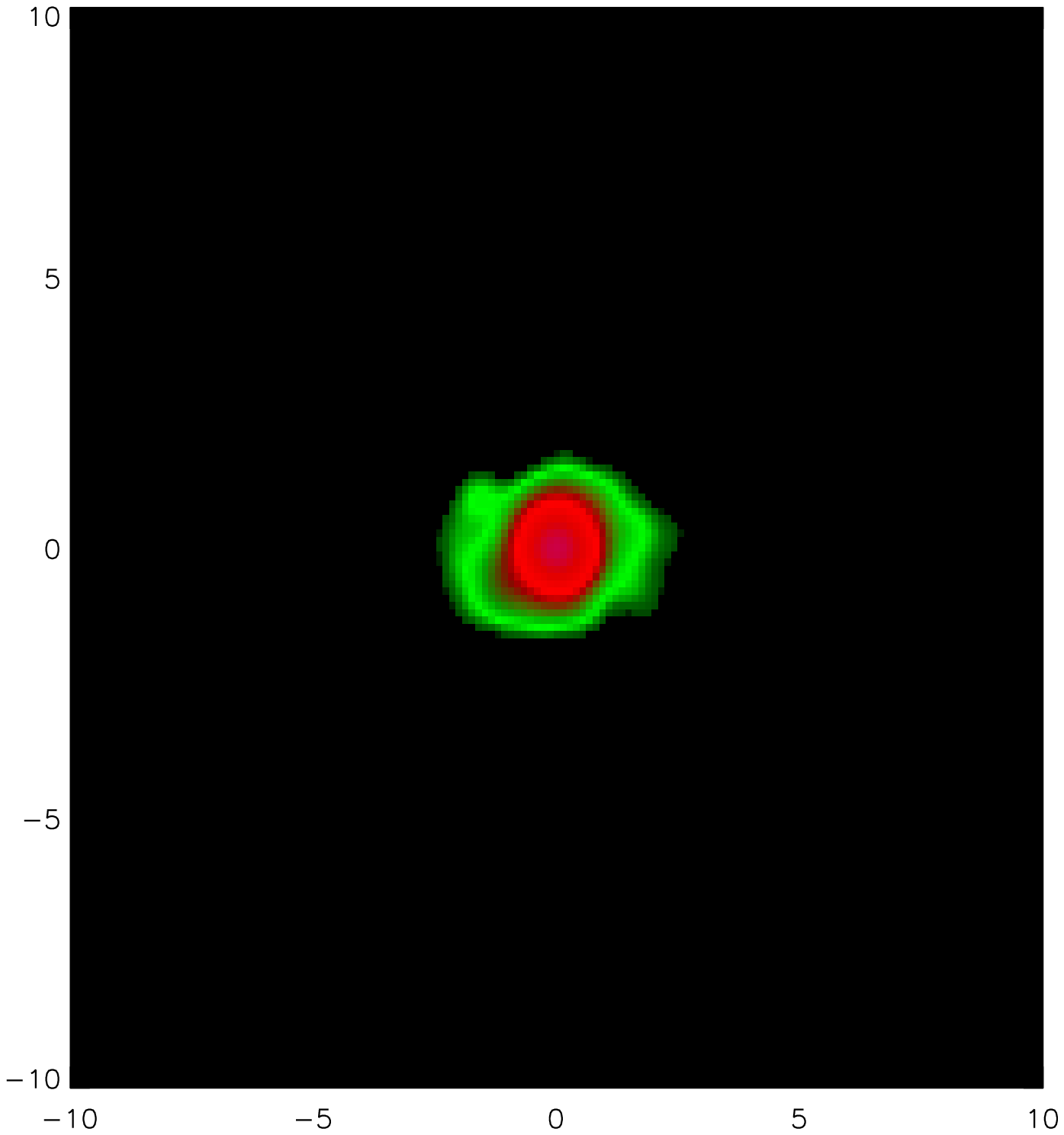,height=2.in,width=2.in}}}
\end{center}
\caption{(a) SB10 at $t=1.165 h^{-1} \rm Gyr$, (b) SB10 at $t=1.205 h^{-1} \rm Gyr$ (c) SB10 at $t=1.330 h^{-1} \rm Gyr$}
\end{figure}

\begin{figure}[h] \begin{center}
\centerline{\psfig{file=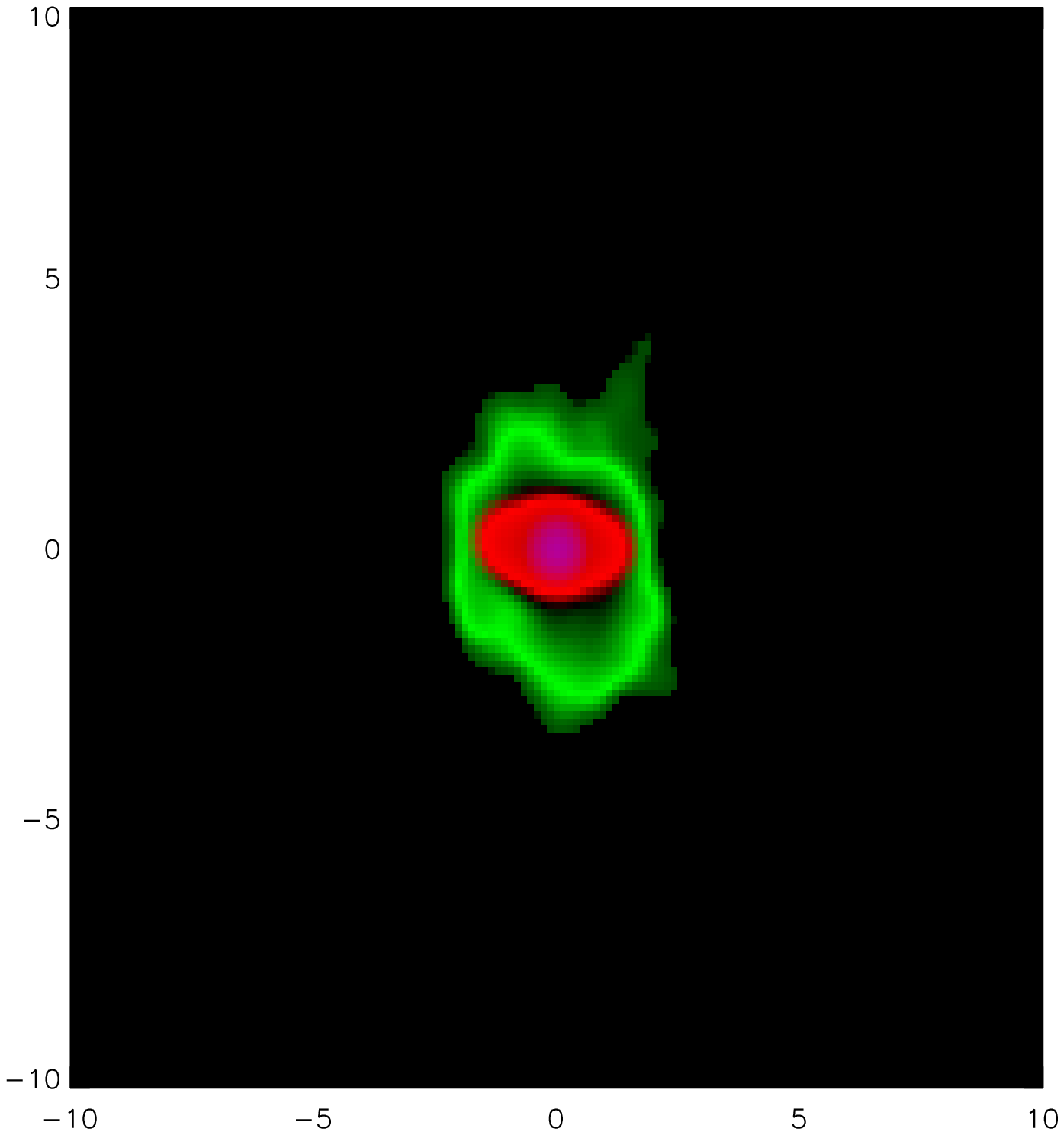,height=2.in,width=2.in}
\psfig{file=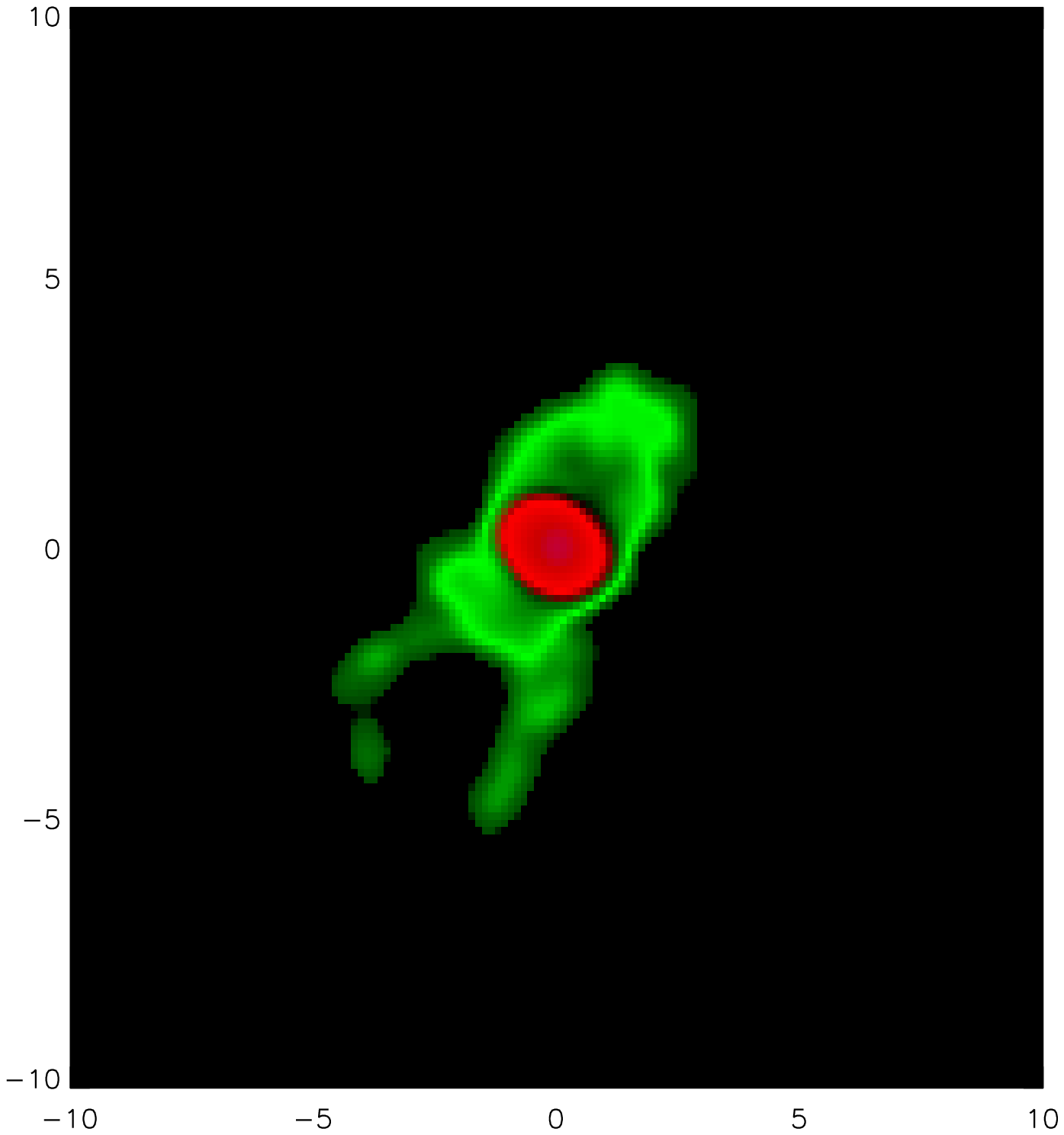,height=2.in,width=2.in}
{\psfig{file=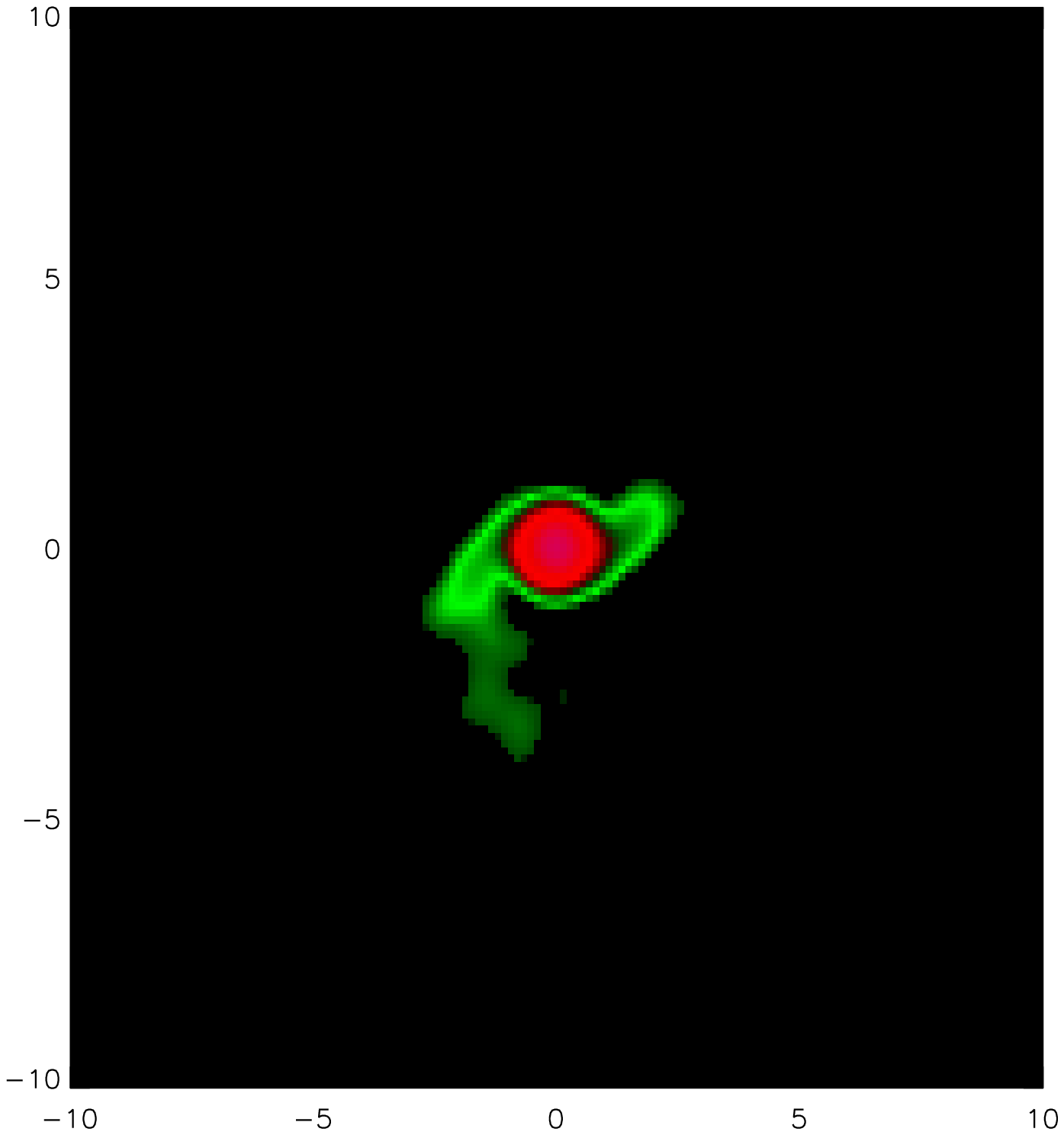,height=2.in,width=2.in}}}
\end{center}
\caption{(a) SB9 at $t=1.165 h^{-1} \rm Gyr$, (b) SB9 at $t=1.205 h^{-1} \rm Gyr$ (c) SB9 at $t=1.330 h^{-1} \rm Gyr$}
\end{figure}

\begin{figure}[h] \begin{center}
\centerline{\psfig{file=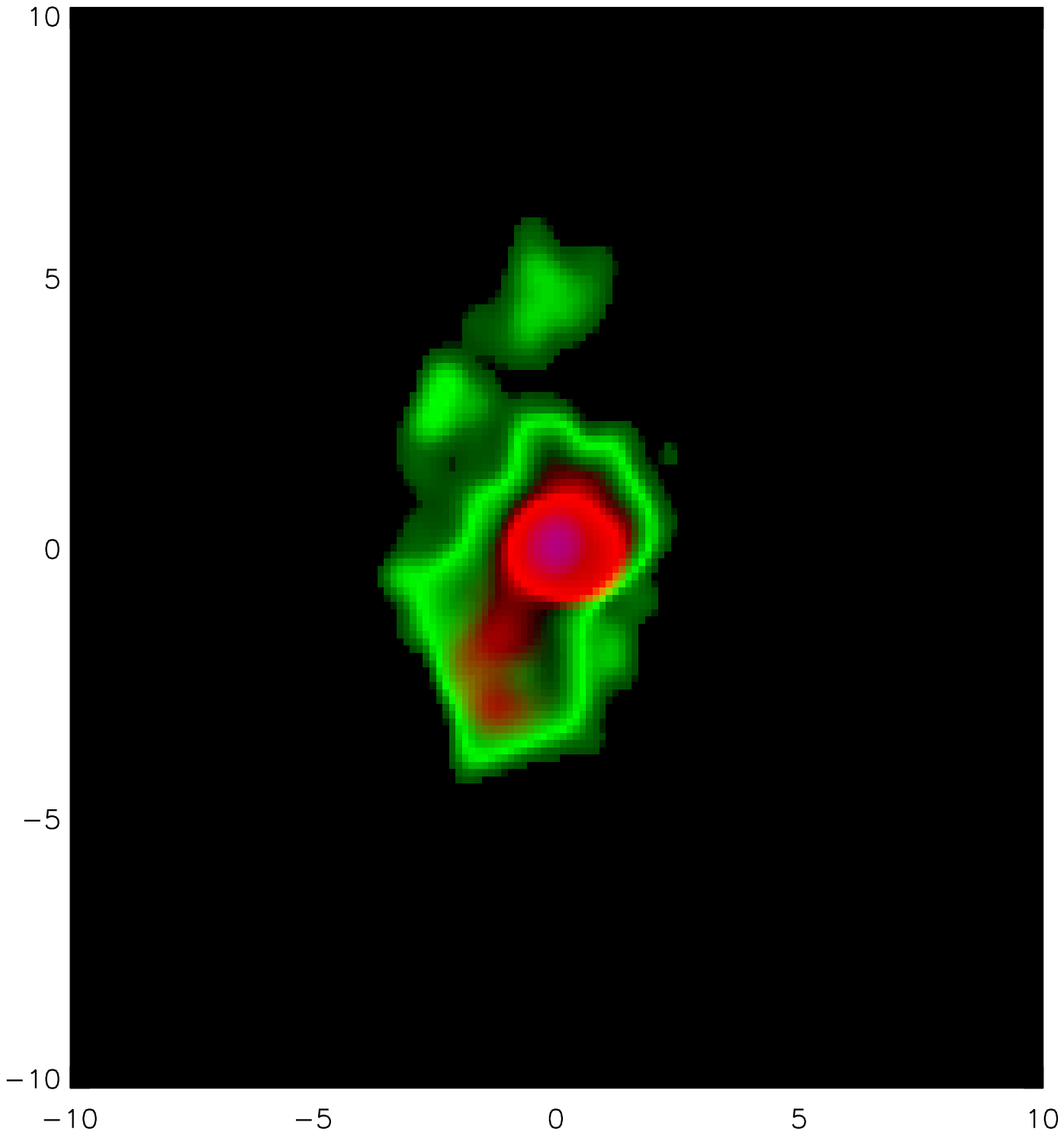,height=2.in,width=2.in}
\psfig{file=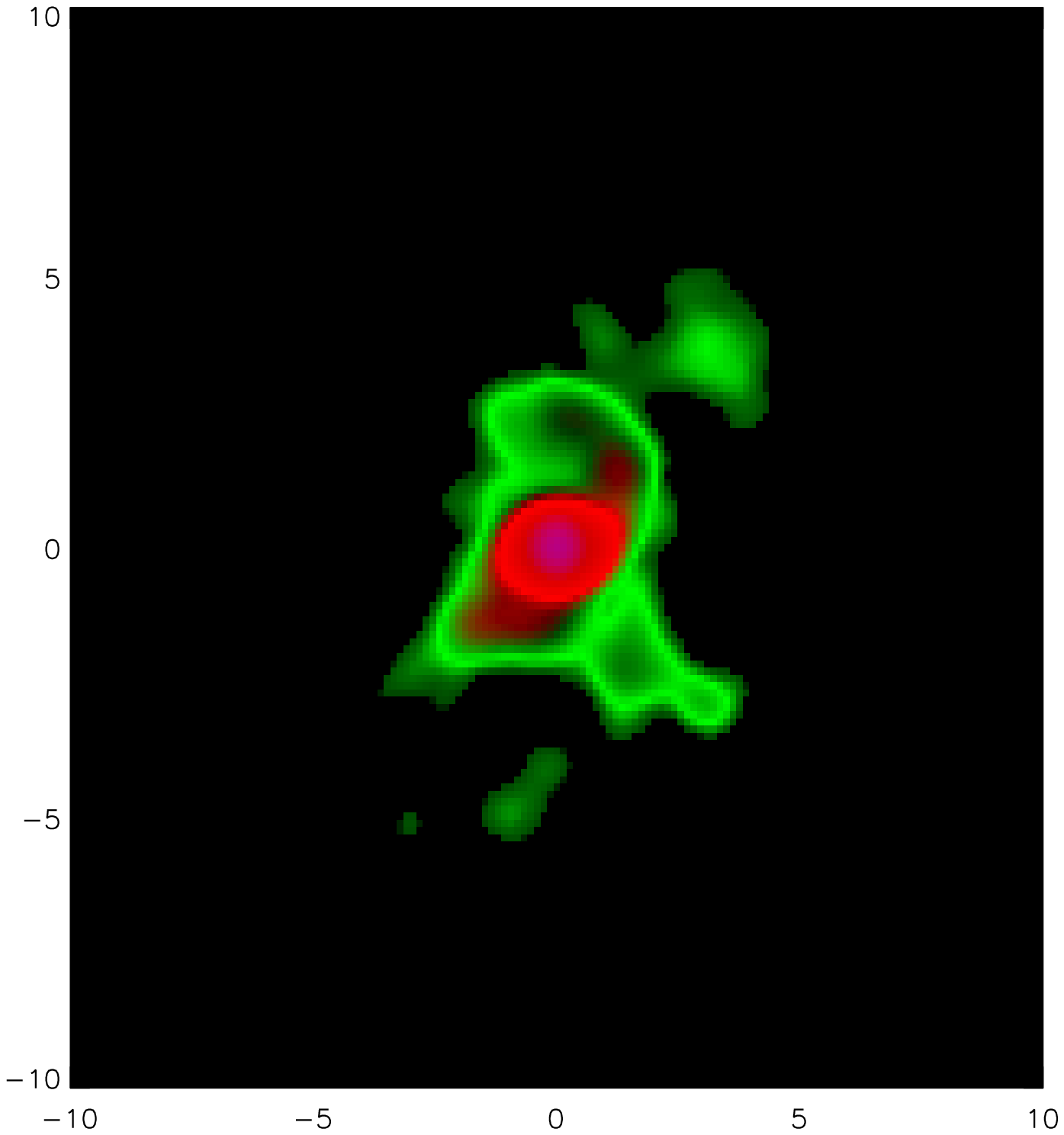,height=2.in,width=2.in}
{\psfig{file=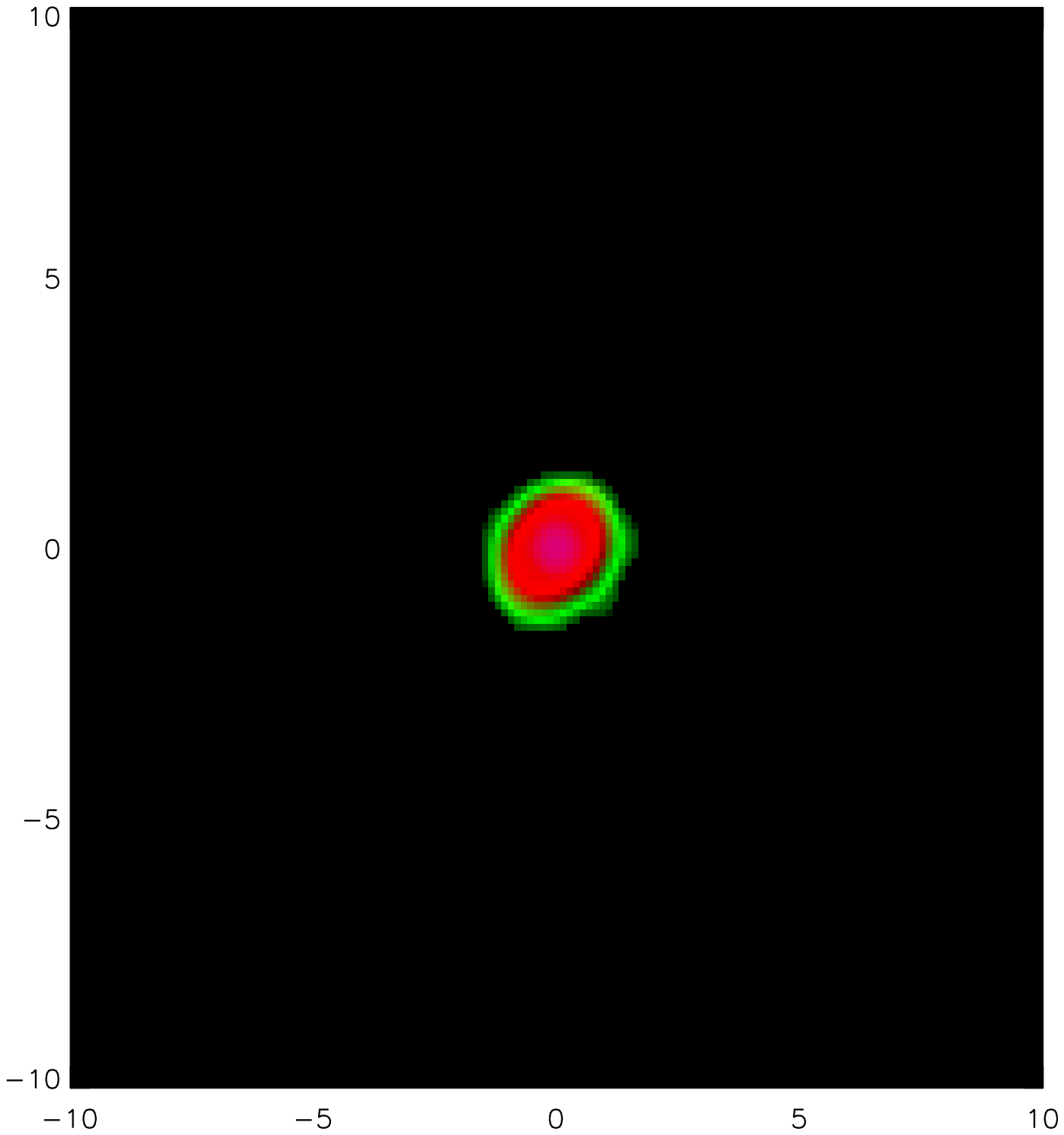,height=2.in,width=2.in}}}
\end{center}
\caption{(a) SB1 at $t=1.165 h^{-1} \rm Gyr$, (b) SB1 at $t=1.205 h^{-1} \rm Gyr$ (c) SB1 at $t=1.330 h^{-1} \rm Gyr$}
\end{figure}

\begin{figure}[h] \begin{center}
\centerline{\psfig{file=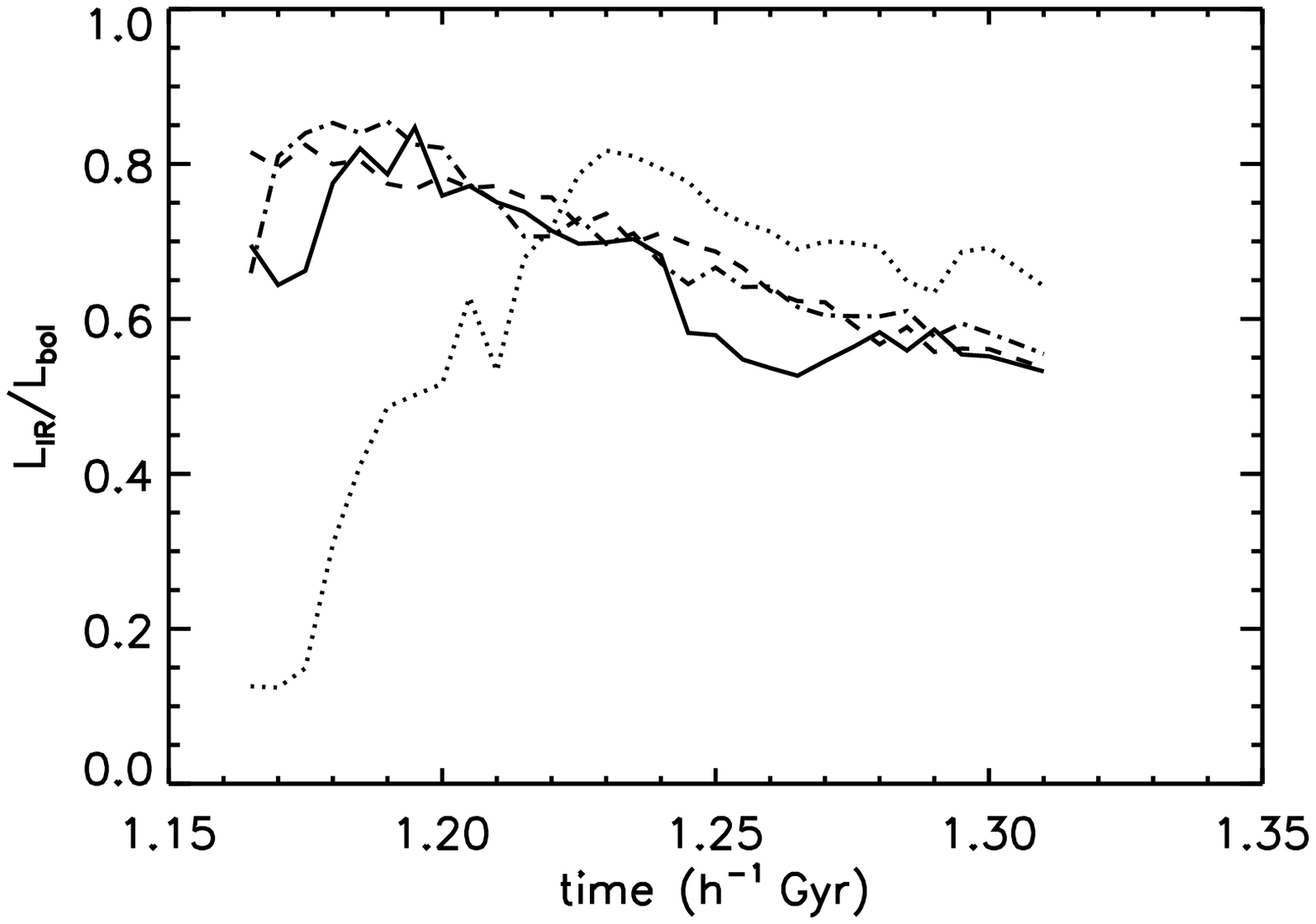,height=2.5in,width=2.5in}
\psfig{file=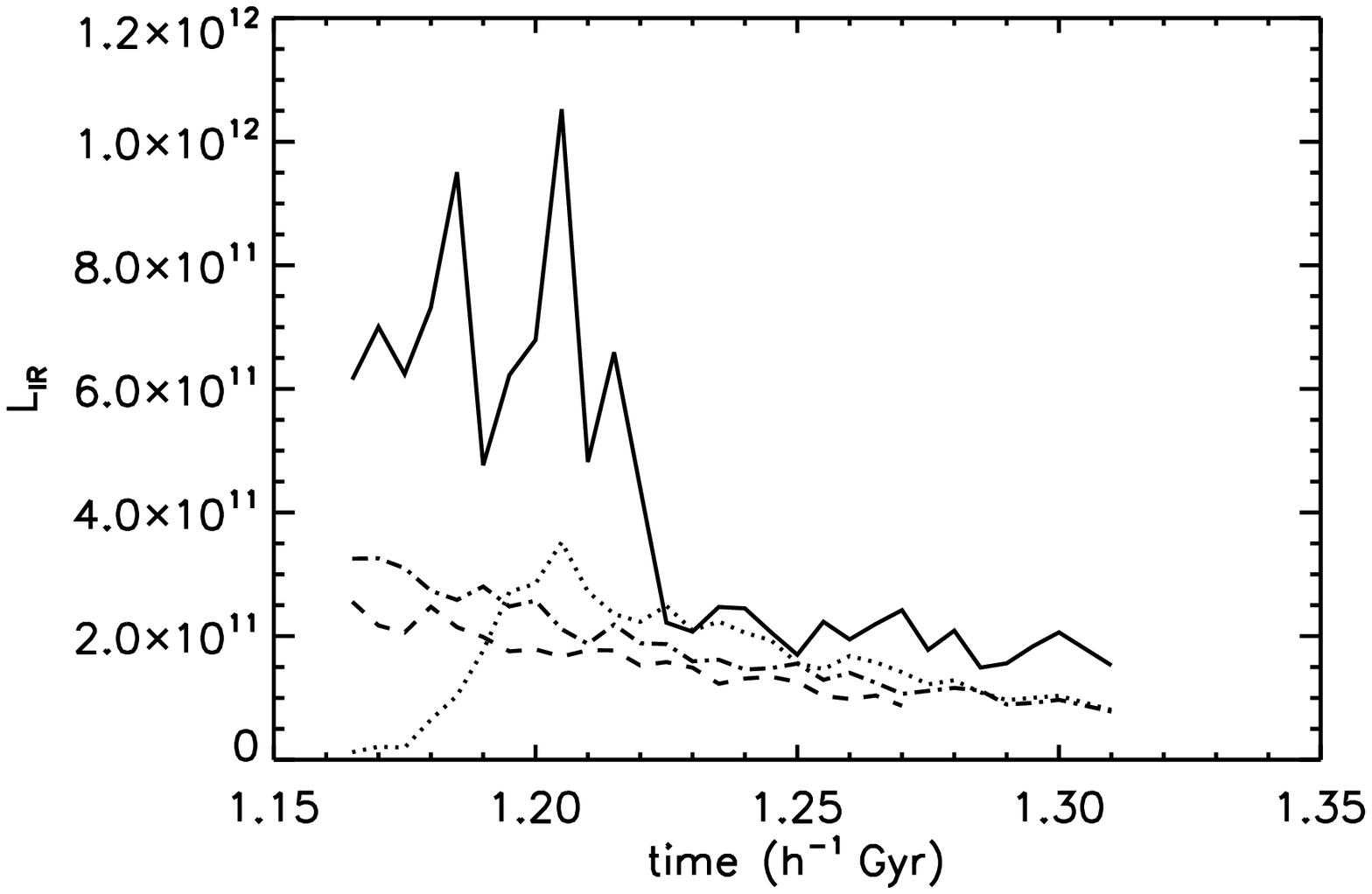,height=2.5in,width=2.5in}
\psfig{file=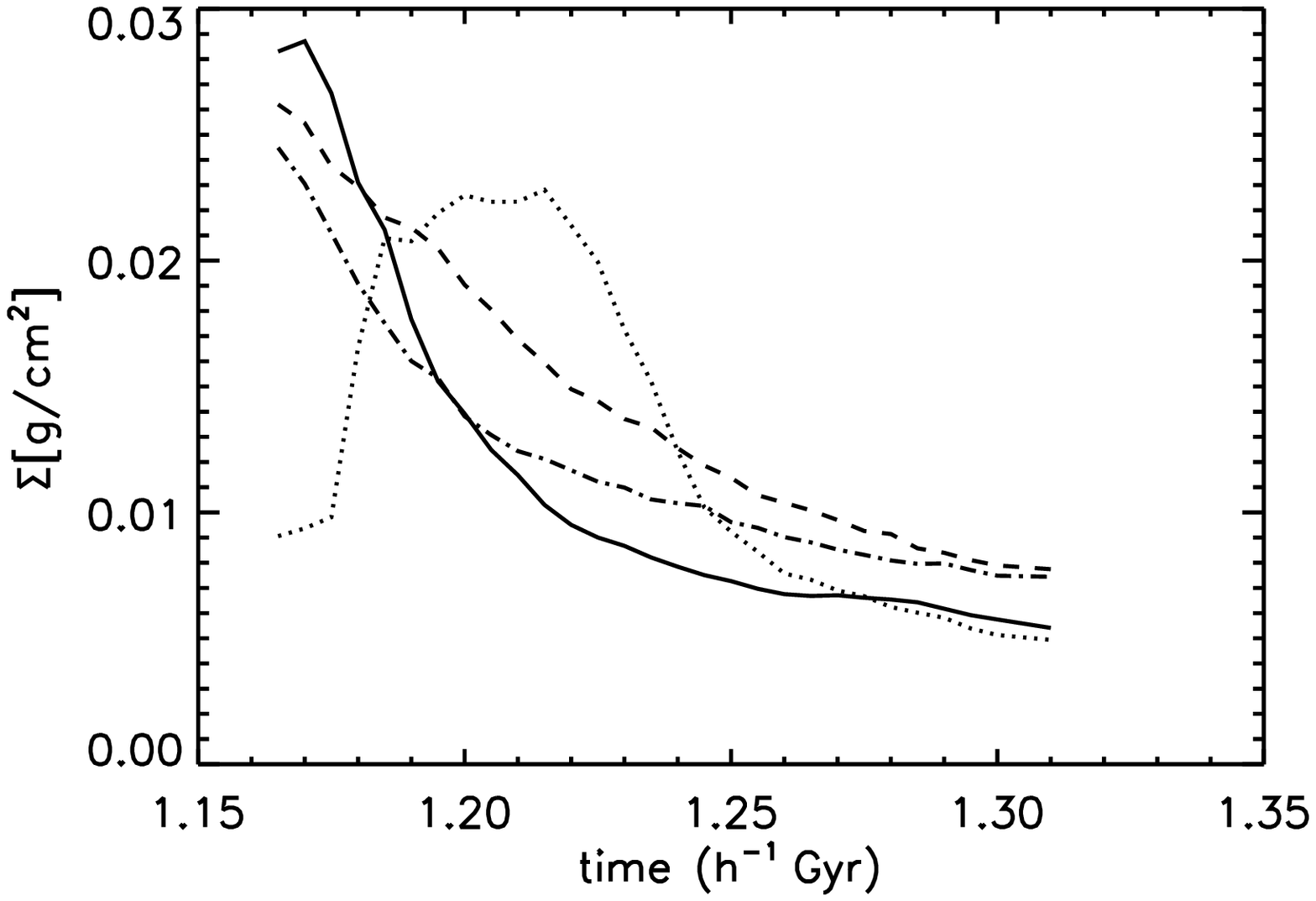,height=2.5in,width=2.5in}}
\end{center}
\caption{(a)Evolution of infrared to bolometric luminosities from simulation with AGN feedback(solid line), SB1 (dashed line), SB9 (dash-dotted line), and SB10 (dotted line),(b) Time variation of infrared luminosity, (c) Time
variation of gas surface density}
\end{figure}

\begin{figure}[h] \begin{center}
\centerline{\psfig{file=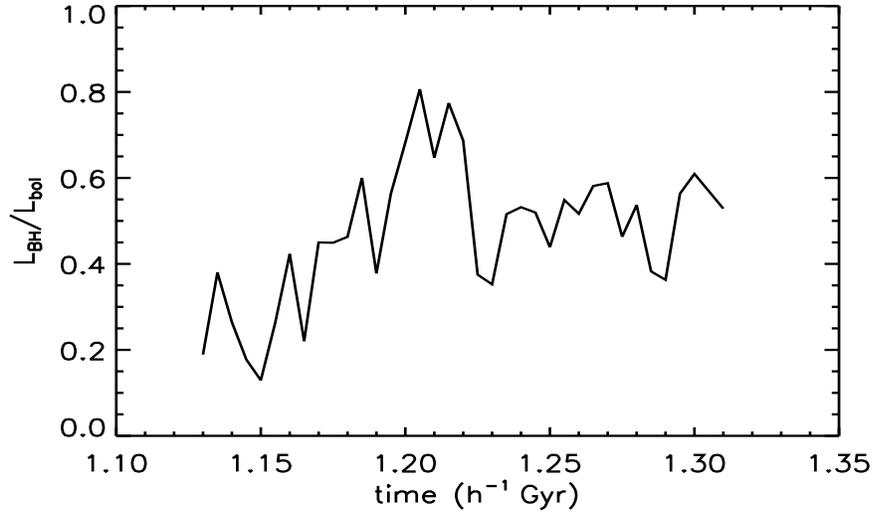,height=3.in,width=5.in}}
\end{center}
\caption{Ratio of black hole luminosity to the total bolometric luminosity (which is the sum of the luminosity from the black hole and the stars) as a function of time.  The AGN dominates in its contribution to the total luminosity for only a short period of time during the active phase, $t\sim 1.2 h^{-1} \rm Gyr$}
\end{figure}

\begin{figure}[h] \begin{center}
\centerline{\psfig{file=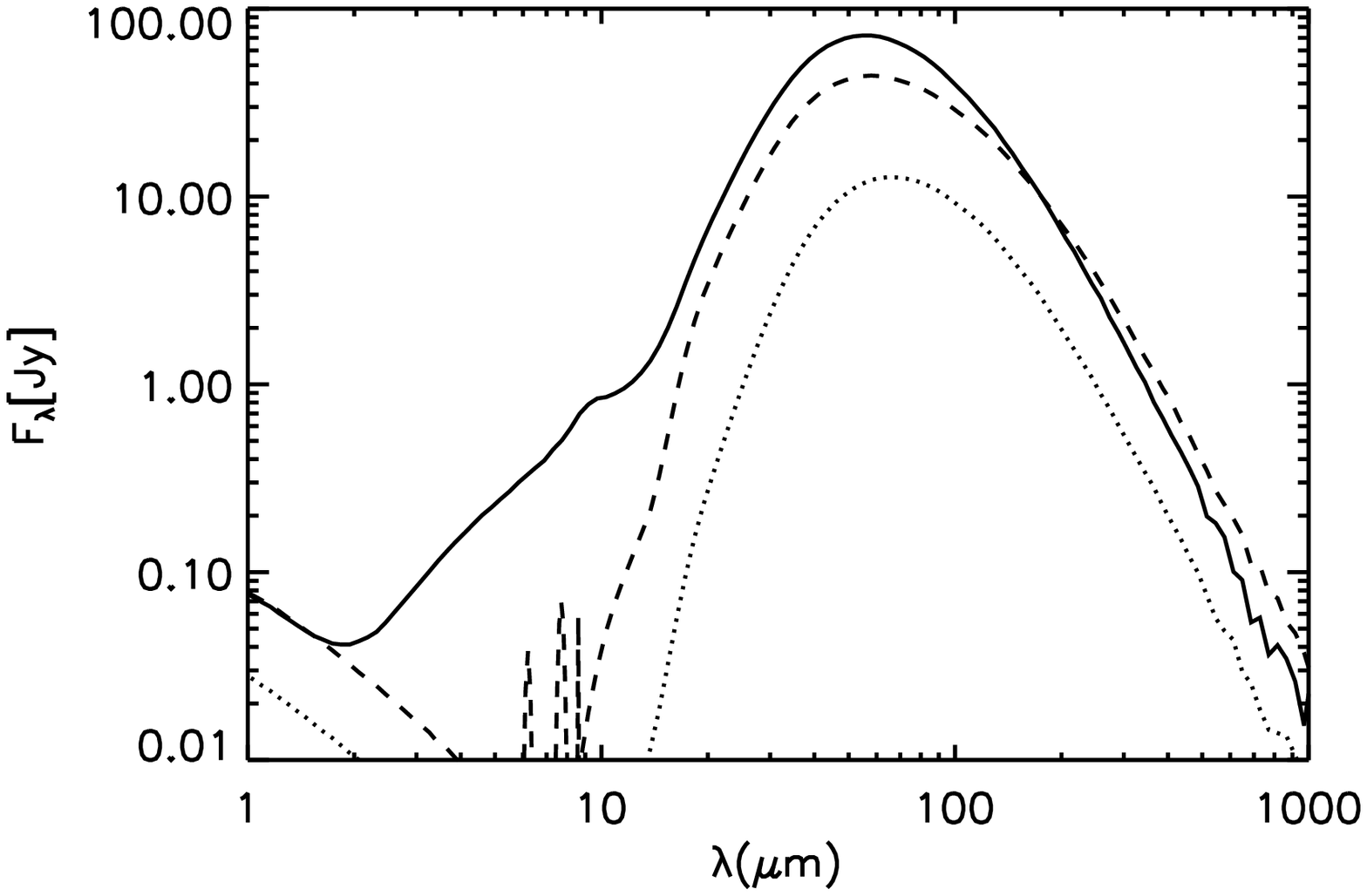,height=2.in,width=2.in}
\psfig{file=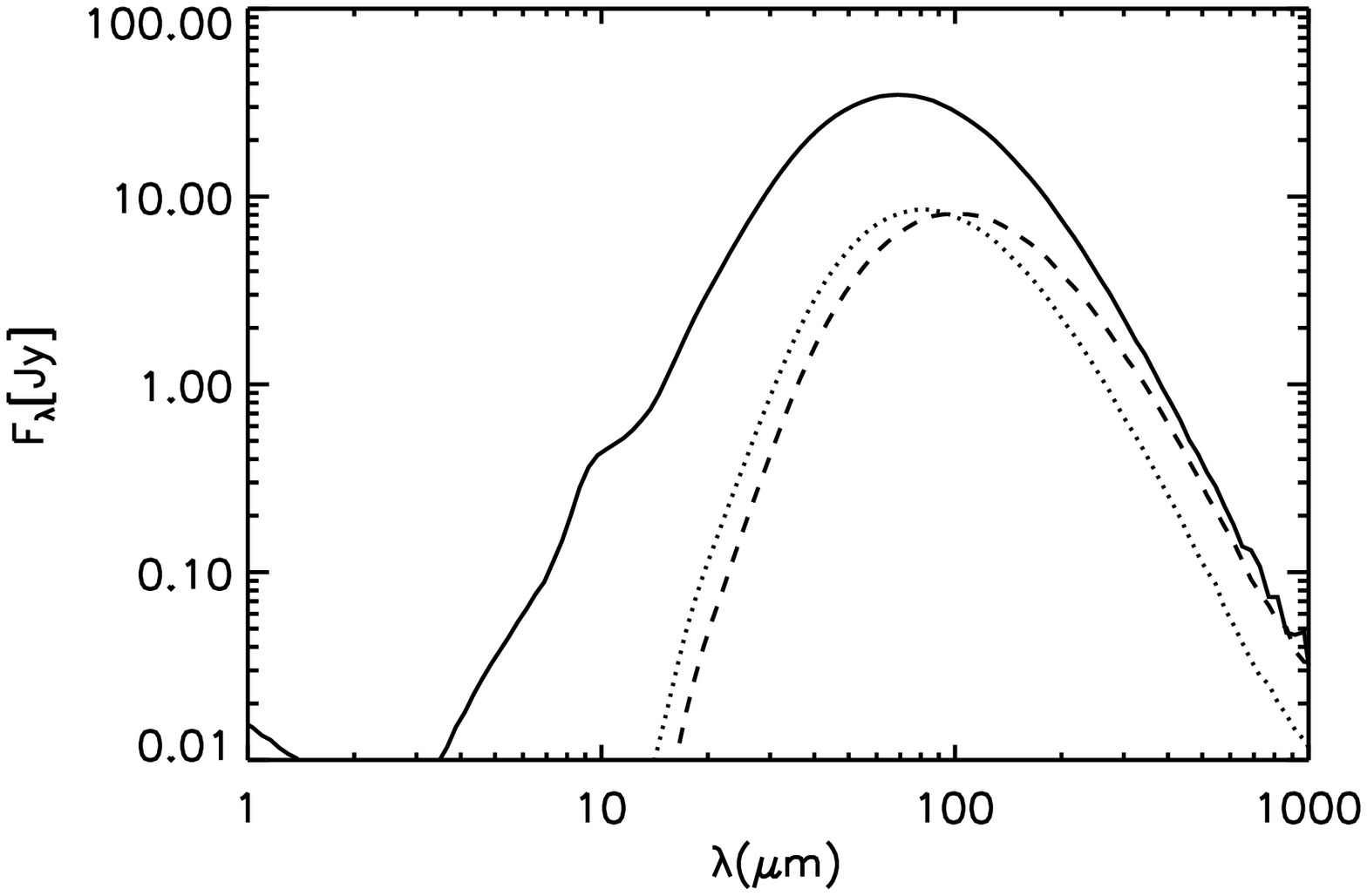,height=2.in,width=2.in}
\psfig{file=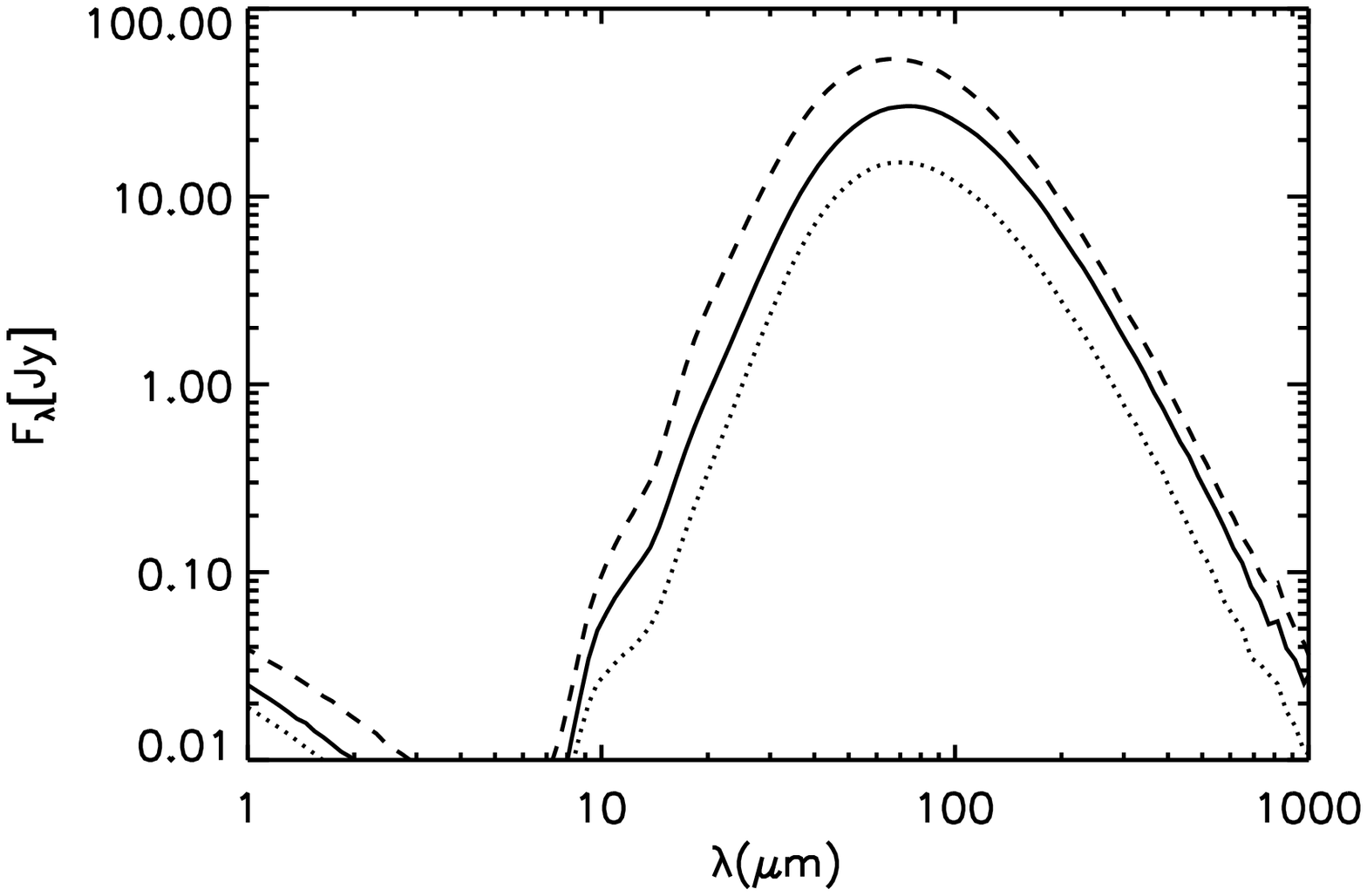,height=2.in,width=2.in}
\psfig{file=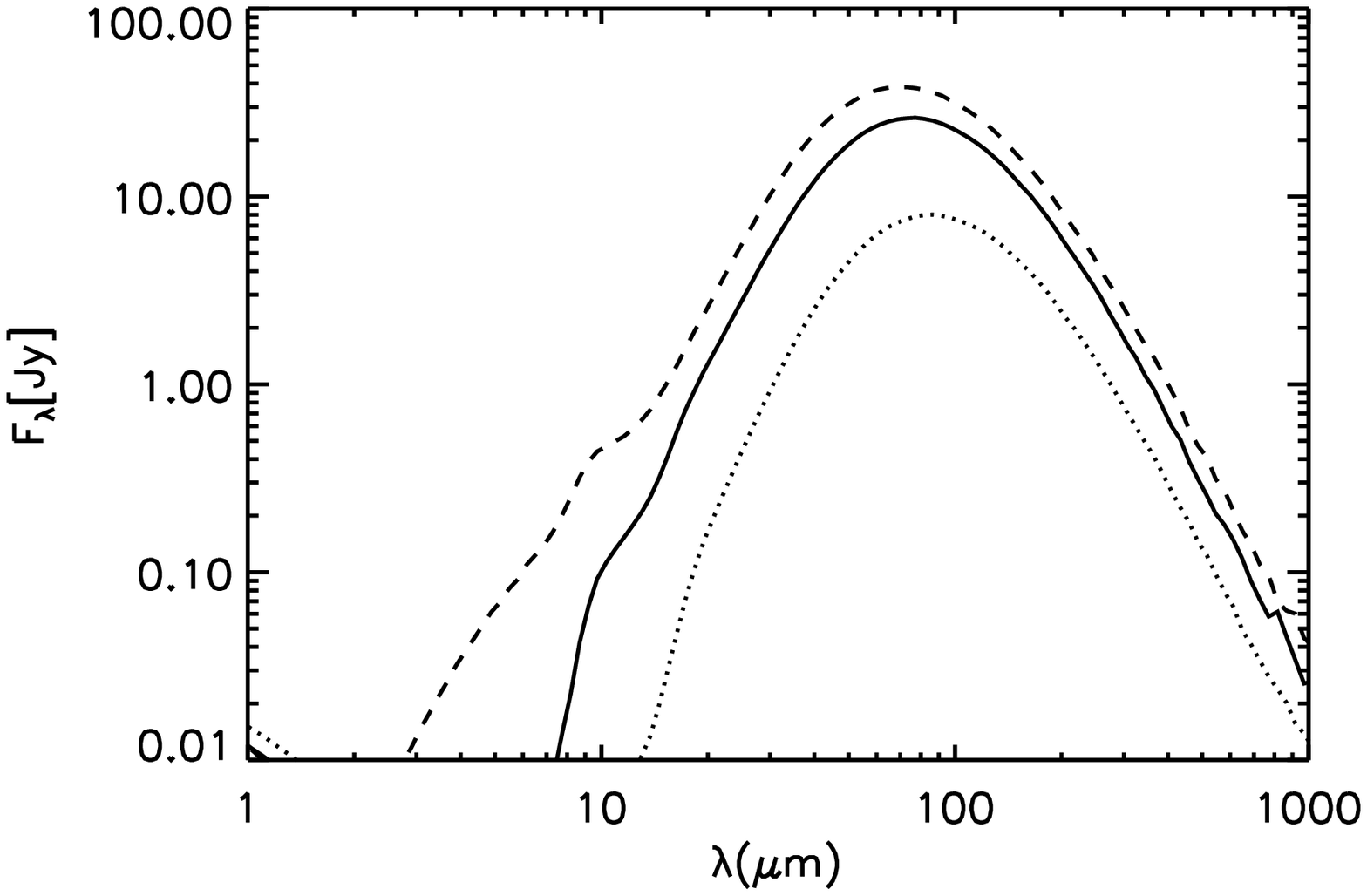,height=2.in,width=2.in}}
\end{center}
\caption{(a) Evolution of SEDs from AGN feedback simulation, $t=1.165 h^{-1} ~\rm Gyr$  (dashed line), $t=1.205 h^{-1}~ \rm Gyr$ (solid line), $t=1.330 h^{-1} ~\rm Gyr$ (dotted line), (b) starburst driven feedback simulation SB10, (c) starburst driven feedback simulation SB9, (d) starburst driven feedback simulation SB1 (all are shown at the same times)}
\end{figure}

\begin{figure}[h] \begin{center}
\centerline{\psfig{file=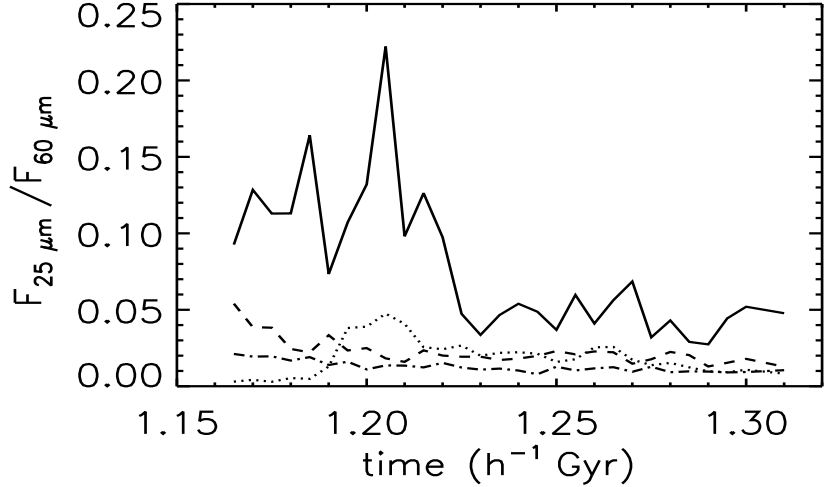,height=3.in,width=5.in}}
\end{center}
\caption{Evolution of $F_{\lambda}(25~\micron)/F_{\lambda}(60~\micron)$ colors from simulation with AGN feedback (solid line), SB1 (dash-dotted line), SB9 (dashed line), and SB10 (dotted line). Note that the warm phase, $t \sim 1.2 \rm h^{-1}~Gyr$, correlates with the peak in relative contribution from the AGN to the total luminosity (shown in Figure 6) and the sharp decline in the gas surface density shown in Figure 5c. }
\end{figure}

\begin{figure}[h] \begin{center}
\centerline{\psfig{file=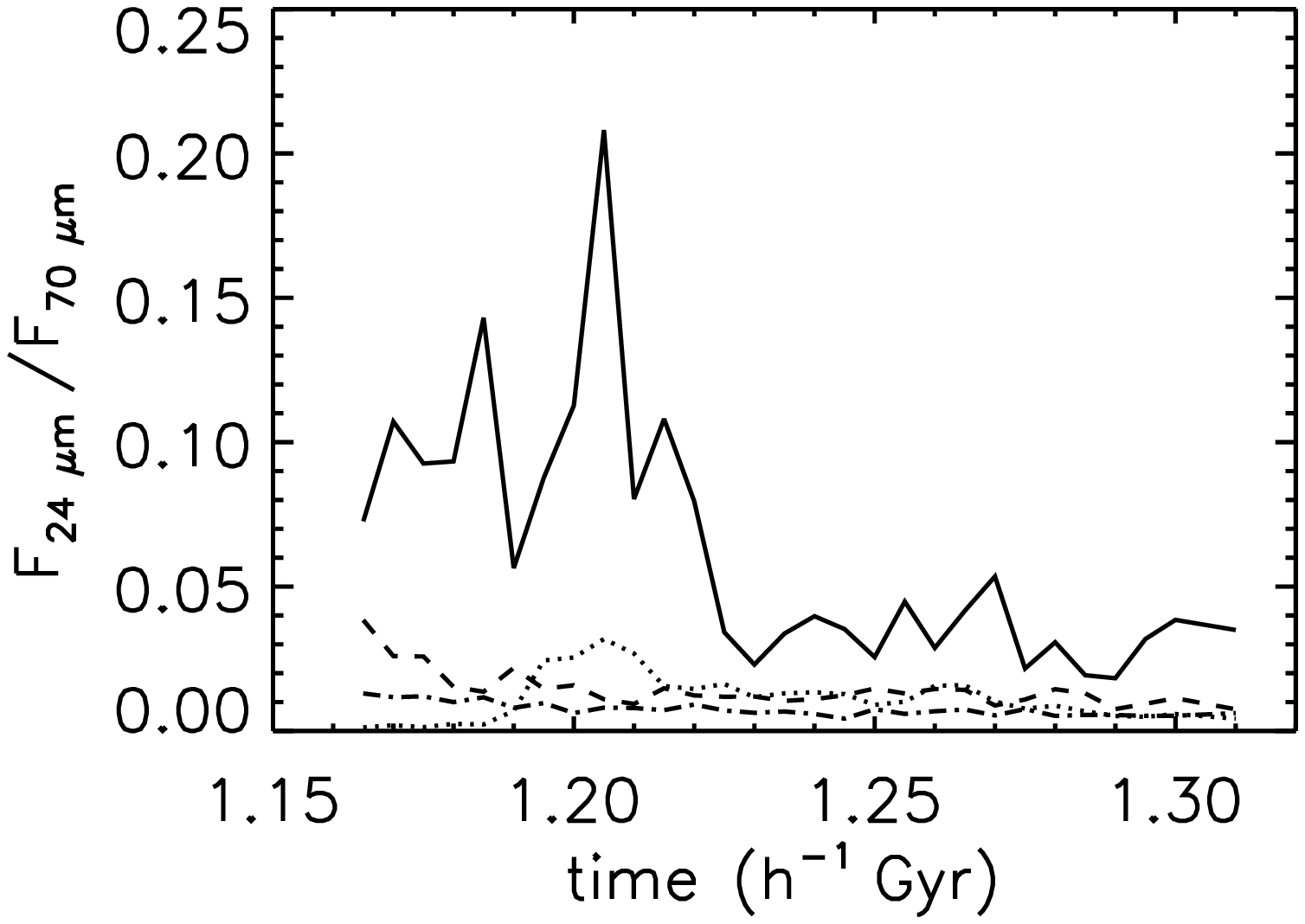,height=3.in,width=3.5in}
\psfig{file=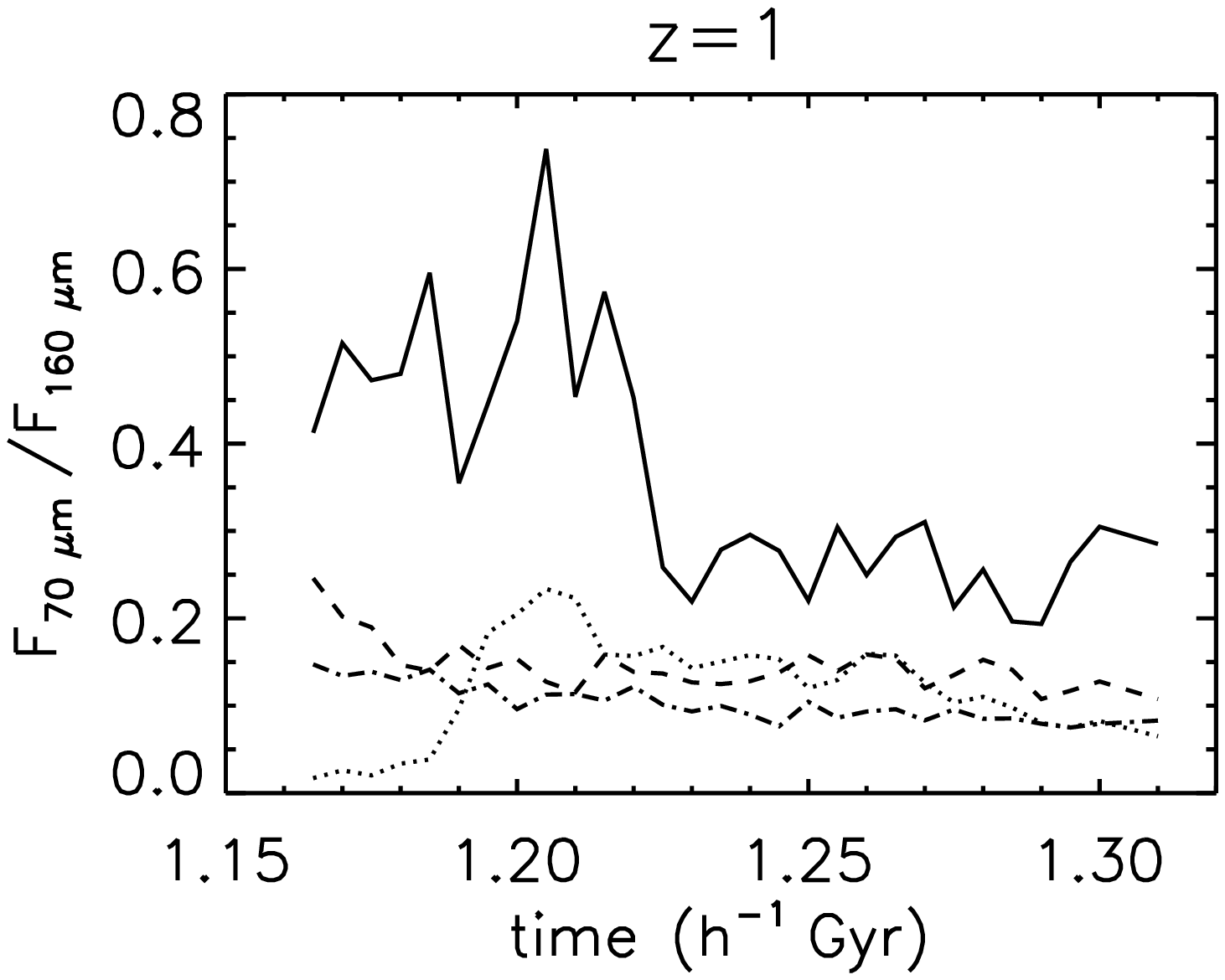,height=3.in,width=3.5in}}
\end{center}
\caption{The Cold-Warm Trend in Spitzer's MIPS Bands: (a) for local galaxies in observed frame $F_{\lambda}(24~\micron)/F_{\lambda}(70~\micron)$, (b) at $z=1$ in observed frame $F_{\lambda}(70~\micron)/F_{\lambda}(160~\micron)$.  In observed $F_{\lambda}(70~\micron)/F_{\lambda}(160~\micron)$ for $z=1$ systems, ``warm'' corresponds to $F_{\lambda}(70~\micron)/F_{\lambda}(160~\micron) \ga 0.6$.}
\end{figure}

\begin{figure}[h] \begin{center}
\centerline{\psfig{file=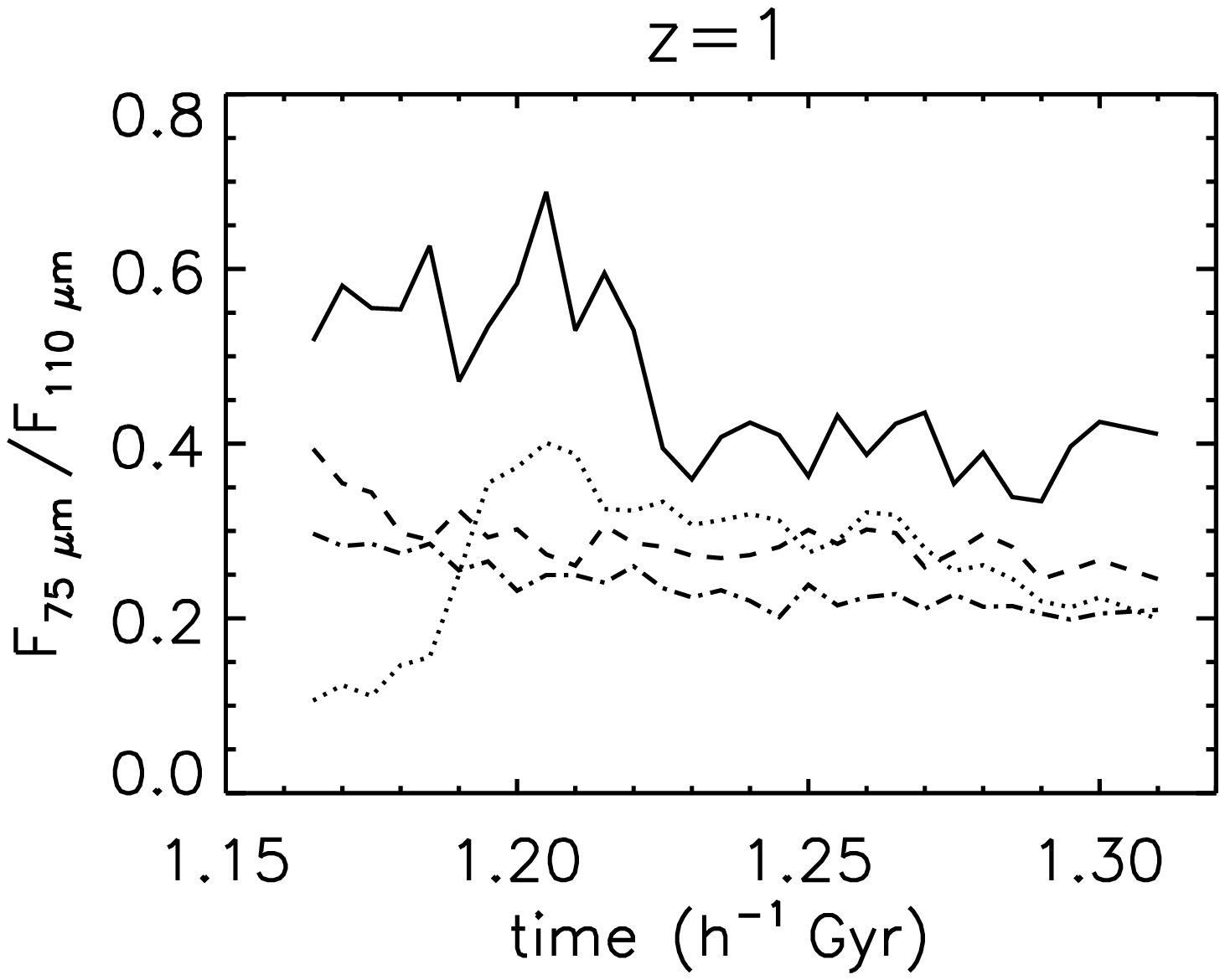,height=3.in,width=3.5in}
\psfig{file=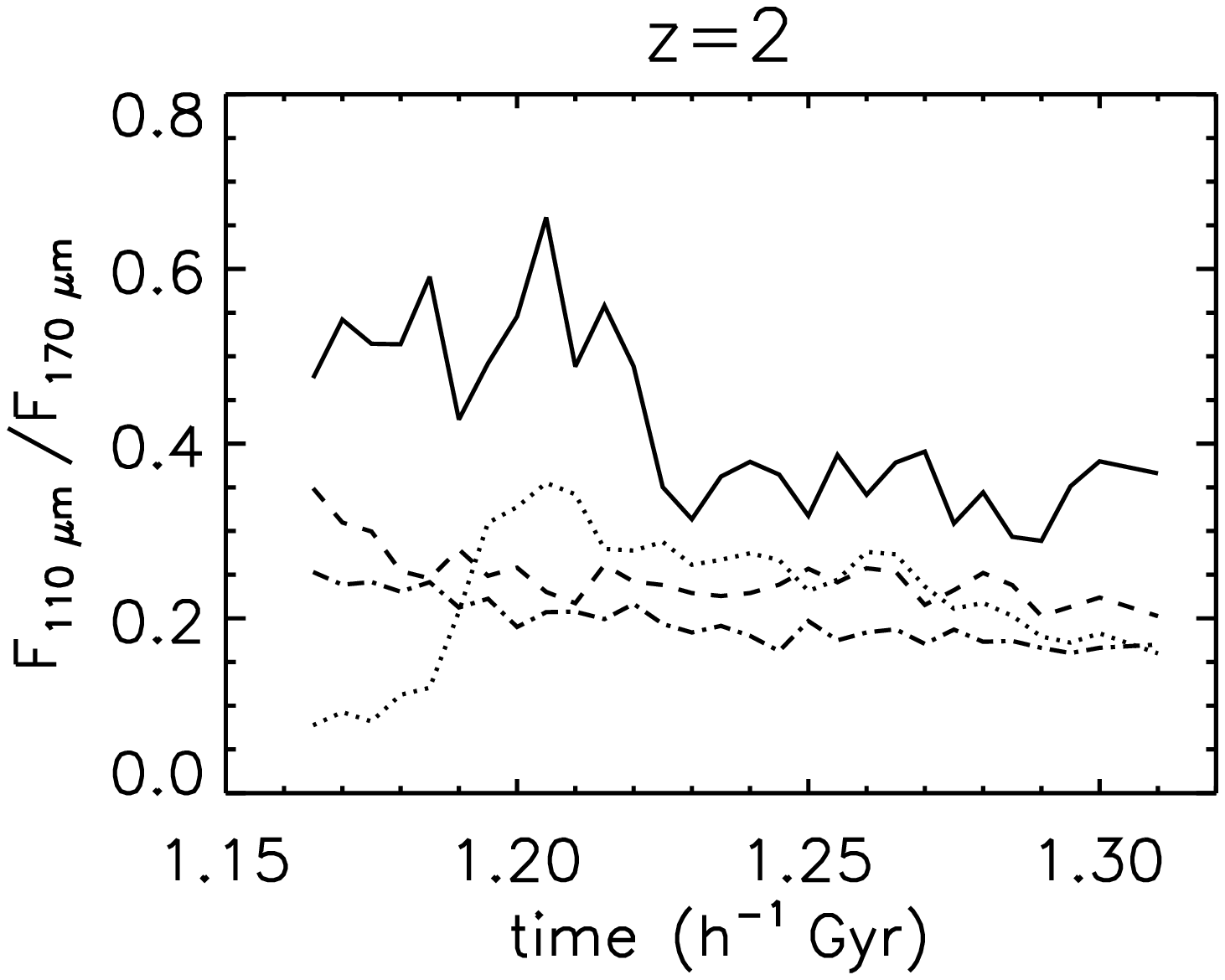,height=3.in,width=3.5in}}
\end{center}
\caption{The Cold-Warm Trend in Herschel's PACS Bands: (a) at $z=1$ in observed frame $F_{\lambda}(75~\micron)/F_{\lambda}(110~\micron)$, here ``warm'' for $z=1$ corresponds to $F_{\lambda}(75~\micron)/F_{\lambda}(110~\micron) \ga 0.6$.(b) At $z=2$ in observed frame $F_{\lambda}(110~\micron)/F_{\lambda}(170~\micron)$, ``warm'' means $F_{\lambda}(110~\micron)/F_{\lambda}(170~\micron) \ga 0.6$.}
\end{figure}

\begin{figure}[h] \begin{center}
\centerline{\psfig{file=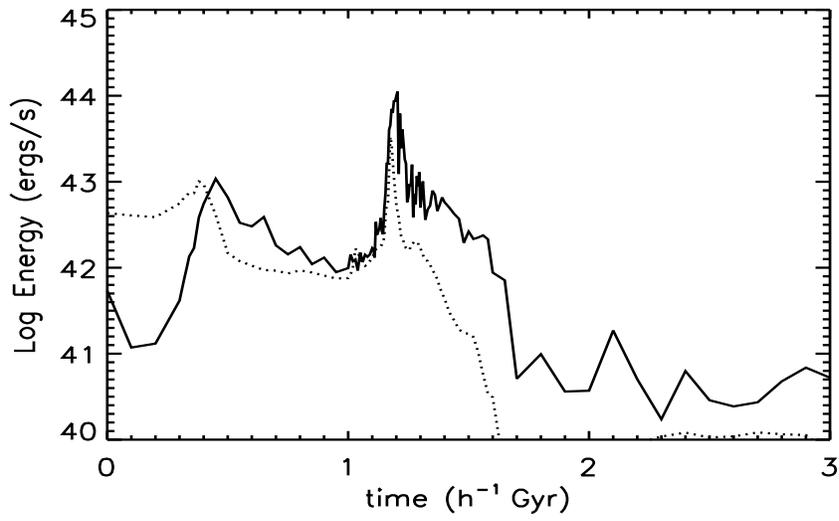,height=3.in,width=5.in}}
\end{center}
\caption{Energy injected into ISM vs. time from AGN feedback (solid line) and 
starburst feedback (dotted line)}
\end{figure}


\clearpage

\begin{references}
\reference{} Aguirre, A., et al. 2001, ApJ, 560, 599
\reference{} Aguirre, A., et al. 2001, ApJ, 561, 521
\reference{} Barnes, J.E. \& Hernquist, L. 1991, ApJ, 370, L65
\reference{} Barnes, J.E. \& Hernquist, L. 1996, ApJ, 471, 115
\reference{} Blitz, L., Fukui, Y., et al., 2006, to appear in Protostars \& Planets V., astro-ph/0602600
\reference{} Brandl, B.R., et al., 2004, ApJS, 154, 188 
\reference{} Bruzual, G.A., \& Charlot, S., 1993, ApJ, 405, 538
\reference{} Chakrabarti, S. \& McKee, C.F., 2005, ApJ, 631,792 [CM05]
\reference{} Chakrabarti, S., \& McKee, C.F., in preparation
\reference{} Chakrabarti, S., \& Whitney, B.A., in preparation
\reference{} Chakrabarti,S., et al. 2006b, submitted to ApJ, astro-ph/0610860
\reference{} Charmandaris, V., et al. 2002, A\&A, 391, 429
\reference{} Cox, T.J., et al. 2006a, ApJ, 643, 692
\reference{} Cox, T.J., et al. 2006b, in prep
\reference{} De Buizer, J.M., Osorio, M., \& Calvet, N., 2005, ApJ, 635, 452 
\reference{} De Grijp et al. 1985, Nature, 314, 240
\reference{} Di Matteo, T., Springel, V., \& Hernquist, L., 2005, Nature, 433, 604
\reference{} Downes, D. \& Solomon, P.M. 1998, ApJ, 507, 615
\reference{} Draine, B.T. \& Lee, H.M., 1984, ApJ, 285, 89
\reference{} Dunne, L., Eales., S. et al., 2000, MNRAS, 315, 115
\reference{} Dunne, L., \& Eales, S.A., 2001, MNRAS, 327, 697
\reference{} Efstathiou, G., \& Rowan-Robinson, M., 1995, MNRAS, 273, 649
\reference{} Efstathiou, G. et al. 2000, MNRAS, 313, 734
\reference{} Egami, E. et al. 2004, ApJS, 154, 130
\reference{} Elvis, M., et al., 1994, ApJS, 95,1
\reference{} Erb, D.K., Shapley, A.E., et al., 2006, astro-ph/0602473
\reference{} Farrah, D. et al. 2002, MNRAS, 335, 1163
\reference{} Farrah, D. et al. 2003, MNRAS, 343, 585
\reference{} Farrah, D. et al. 2005, ApJ, 626, 70
\reference{} Finlator, K. et al. 2006, ApJ, 639, 672F
\reference{} George, I.M., Turner, T.J., et al. 1998, ApJS, 114, 73
\reference{} Goldader, J.D., Meurer, G., Heckman, T.M, et al. 2002, ApJ, 568, 651
\reference{} Granato et al. 2000, ApJ, 542, 710
\reference{} Guhathakurta, P., \& Draine, B.T., 1989, ApJ, 345, 230
\reference{} Hernquist, L, 1990, ApJ, 356, 359
\reference{} Hopkins, P., et al. 2005a, ApJL, 625, L71
\reference{} Hopkins, P., et al. 2005b, ApJ, 630, 716
\reference{} Hopkins, P., et al. 2006a, ApJS, 163, 1
\reference{} Hopkins, P., et al. 2006b, ApJS, 163, 50
\reference{} Hopkins, P., et al. 2006c,  ApJ, submitted, astro-ph/0602290
\reference{} Kauffmann, G., \& Charlot, S., 1998, MNRAS, 297, 23
\reference{} Klaas, U., Haas, M., et al., 2001, A\&A, 379, 823
\reference{} Krumholz, M.R., \& McKee, C.F., 2005, ApJ, 630, 250
\reference{} Lutz, D., Spoon, H.W.W., et al., 1998, ApJ, 505, 103
\reference{} Lutz, D. et al. 2005, ApJ, 632, 13
\reference{} Magdziarz, P., \& Zdziarski, A.A., 1995, MNRAS, 273, 837
\reference{} Manske, V., \& Henning, Th., 1998, A\&A, 337, 85
\reference{} Marconi, A., \& Hunt, L., 2003, ApJ, 589, L21
\reference{} Marconi, A., Risaliti, G., et al. 2004, MNRAS, 351, 169
\reference{} Mihos, J.C. \& Hernquist, L. 1994, ApJ, 431, L9
\reference{} Mihos, J.C. \& Hernquist, L. 1996, ApJ, 464, 641
\reference{} Mueller, K.E., Shirley, Y.L., et al. 2002, ApJS, 143, 469
\reference{} Nagamine, K. et al. 2005a, ApJ, 618, 23
\reference{} Nagamine, K. et al. 2005b, ApJ, 627, 608
\reference{} Narayanan, D. et al. 2006, ApJ, 642L, 107N
\reference{} Natta, A., \& Panagia, N., 1984, ApJ, 287, 228
\reference{} Night, C. et al. 2006, MNRAS, 366, 705
\reference{} Perola, G.C., et al., 2002, A\&A, 389, 802
\reference{} Purcell, E.M., 1976, ApJ, 206, 685
\reference{} Recchi, S. \& Matteucci, F., 2002, A\&A 384, 799
\reference{} Rigopoulou, D., Spoon, H.W.W., et al. 1999, AJ, 118, 2625
\reference{} Robertson, B., et al., 2004, ApJ, 606, 32
\reference{} Rosolowsky, E., \& Blitz, L, 2005, ApJ, 623, 826
\reference{} Rupke, D.S., Veilleux, S., \& Sanders, D.B., 2005a, ApJ, 632, 751
\reference{} Rupke, D.S., Veilleux, S., \& Sanders, D.B., 2005b, ApJS, 160, 115
\reference{} Sanders, D.B., Soifer, B.T., Elias, J.H., Neugebauer, G., \& 
Matthews, K., 1988, ApJ, 328, L35
\reference{} Scoville, N.Z., Evans, A.S., et al. 1998, ApJ, 492, L107
\reference{} Scoville, N.Z., Evans, N.S., et al., 2000, AJ, 119, 991
\reference{} Siebenmorgen, R., Kruegel, E., \& Mathis, J.S., 1992, A\& A, 266, 501
\reference{} Silva et al. 1998, ApJ, 509, 103
\reference{} Soifer, B.T. et al. 1984, ApJ, 278, L71
\reference{} Soifer, B.T. et al. 1987, ARA\&A, 25, 187
\reference{} Soifer, B.T., et al., 1999, ApJ, 513, 207
\reference{} Soifer, B.T. et al. 2000, AJ, 119, 509
\reference{} Solomon, P.M., Rivolo, A.R., et al., 1987, ApJ, 319, 730
\reference{} Solomon, P.M., Downes, D., et al. 1997, ApJ, 478, 144
\reference{} Spoon, H.W.W., Moorwood, A.F.M., et al. 2004, A\&A, 414, 873
\reference{} Springel, V., 2005, MNRAS, 364, 1105S
\reference{} Springel, V. \& Hernquist, L., 2002, MNRAS, 333, 649
\reference{} Springel, V. \& Hernquist, L., 2003, MNRAS, 339, 289 [SH03]
\reference{} Springel, V., di Matteo, T., \& Hernquist, L, 2005b, MNRAS, 361, 776
\reference{} Springel, V., di Matteo, T., \& Hernquist, L, 2005a, ApJ, 620, L79
\reference{} Sridharan, T.K., Beuther, H., et al., 2002, ApJ, 566, 931
\reference{} Stone, J.M., Xu, J., \& Mundy, L.G., 1995, Nature, 377, 315
\reference{} Surace, J.A., Sanders, D.B., Evans, A.S., 2000, ApJ, 529, 170
\reference{} Surace, J.A. \& Sanders, D.B., 1999, ApJ, 512, 162
\reference{} Telfer, R.C., et al., 2002, ApJ, 565, 773
\reference{} Tran, Q.D., Lutz, D., et al., 2001, ApJ, 552, 527
\reference{} Ueda, Y., Akiyama, M., et al., 2003, ApJ, 598, 886
\reference{} Verma, A. et al. 2005, SSRv, 119, 355
\reference{} Vignali, C., Brandt, W.N., \& Schneider, D.P., 2003, AJ, 125, 433
\reference{} Weingartner, J., \& Draine, B.T., 2001, ApJ, 548, 296
\reference{} Whitney, B.A., Wood, K., et al., 2003, ApJ, 598, 1079
\reference{} Xu et al. 2001, ApJ, 562, 179
\reference{} Yan, L., Chary, R., Armus, L, et al. 2005, ApJ, 628, 604

\end{references}
\end{document}